\documentclass[pre,10pt,showpacs]{revtex4}

\usepackage{amsmath}    
\usepackage{graphicx}   
\usepackage{verbatim}   
\usepackage{color}      
\usepackage{subfigure}  
\usepackage{hyperref}   
\usepackage{pstricks}
\usepackage{epstopdf}

\newcommand{\rf}[1]{Eq.~(\ref{#1})}

\def\etal{\textit{et al.~}}

\def\e{\textrm{e}}
\newcommand{\tro}[1]{\textrm{Tr}\left\{\rho_0 {#1}\right\}}
\newcommand{\tr}[1]{\textrm{Tr}\left\{{#1}\right\}}

\def\ompl{\omega_\mathrm{pl}}

\usepackage{ifthen}
    \newboolean{showTODO}%
    \newboolean{showdetails}%
    \setboolean{showTODO}{false}%
    \setboolean{showdetails}{false}%

  \ifthenelse{\boolean{showdetails}}
             {
\newenvironment{details}[0]{%

\vspace*{0.02\textwidth}
\indent
\begin{minipage}[t]{0.80\textwidth}
\scriptsize
}{%
\end{minipage}%
\vspace*{0.02\textwidth}

} } {%
\newenvironment{details}[0]{%
\comment%
}{%
\endcomment%
}%
             }%

\begin{document}

\title{
Dielectric function beyond the random-phase approximation: Kinetic
   theory versus linear response theory
}

\author{H. Reinholz}
\affiliation{Universit\"at Rostock, Institut f\"ur Physik, 18051 Rostock, Germany}
\affiliation{Johannes-Kepler-Universit\"at, Institut f\"ur Physik, 4040 Linz, Austria}
\affiliation{University of Western Australia School of Physics, WA 6009 Crawley, Australia}

\author{G. R\"opke}
\affiliation{Universit\"at Rostock, Institut f\"ur Physik, 18051 Rostock, Germany}

\email{heidi.reinholz@uni-rostock.de}

\date{\today}

\begin{abstract}
Calculating the frequency dependent dielectric function for strongly coupled plasmas, the relations  within kinetic theory and linear response theory are derived and discussed in comparison. In this context, we give a proof that the Kohler variational principle can be extended to arbitrary frequencies.  It is shown to be  a special case of the Zubarev method for the construction of a non-equilibrium statistical operator from the principle of the extremum of entropy production.
Within kinetic theory, the commonly used energy dependent relaxation time approach is strictly valid only for the Lorentz plasma in the static case. It is compared with the result from linear response theory that includes  electron-electron interactions 
and applies for arbitrary frequencies, including bremsstrahlung emission. It is shown how a general approach to linear response  encompasses the different approximations and opens options for systematic improvements.

\end{abstract}

\maketitle

\section{Introduction}

Different approaches have been elaborated to evaluate the response of a plasma to external time and space dependent electric fields. 
This applies, e.g., to absorption and  emission of radiation (in particular bremsstrahlung), Thomson scattering, and stopping power.
The dielectric function $\epsilon(\vec k, \omega)$ depending on the wave number $\vec k$ and frequency $\omega$  as the central quantity  
is related to the polarization function, the dynamical conductivity or the dynamical structure factor.  
The random phase approximation (RPA)  is improved if collisions are taken into account. 
In this context, a non-local dynamical collision frequency  is introduced \cite{Berk92,rrw99a,rerrw00,Reinholz05}.
Alternatively, the concept of a local-field factor \cite{Hubbard57} 
can be extended to dynamical local-field corrections \cite {Ichimaru82,Ichi5,Ashcroft94,rrw99a}. 
In the present work, we focus on the dynamical conductivity and restrict ourselves to the long-wavelength limit $k \to 0$, i.e., 
the response of a charged particle system to a homogeneous, time-dependent electrical field.

A well-known expression for the dc conductivity of a fully ionized plasma in the classical, low-density limit has been given 
by Spitzer and H\"arm \cite{Spitzer53} within kinetic theory (KT).
Further approaches by Lee and More \cite{LeeMore},  Stygar \cite{stygar02}, and others improved the electron-ion interaction 
using the relaxation time approach. However, to recover the Spitzer result for the conductivity, 
electron-electron collisions have to be taken into account. This is not consistently possible within the relaxation time approach \cite{Appel61}. 
We discuss a general approach that allows also for a systematic treatment of electron-electron collisions.

The investigation of time-dependent fields is somehow difficult in KT, too. Often, a combination of the collisionless kinetic equation with the relaxation time ansatz is used, 
see Landau and Lifshits \cite{landau10},  Dharma-wardana \cite{dharma06}, or Kurilenko \etal \cite{Berk92,Kuri95}. 
It has been emphasized by Landau and Lifshits \cite{landau10}  that such an approach is  only applicable in the low frequency limit.   
The high-frequency region, where bremsstrahlung  is relevant, has to be treated in another way. 
In this work, we present general expressions applicable to arbitrary frequencies of the external field.

In linear response theory (LRT), the Kubo formula \cite{Kubo66} was considered as a promising approach to 
the dynamical conductivity in dense, strongly interacting systems at arbitrary degeneracy. 
A  generalized approach to non-equilibrium processes has then been given by Zubarev \textit{et al.} \cite{ZMR2} 
that  will be applied here. It relates transport properties to equilibrium correlation functions such as 
current-current or force-force correlation functions. Different methods can be applied to evaluate these 
correlation functions such as numerical simulations, density functional approaches \cite{dharma06,holst08}, 
or analytical expressions derived from perturbation theory \cite{Roep88,Berk93a,rerrw99}. Note that also strict results such as sum rules can be employed 
to construct the dynamical structure factor, see \cite{Tkachenko,Lee,Murillo}.
We will show how consistent approximations are obtained from a general scheme of non-equilibrium statistical physics
and systematic improvements can be given. 
 
In the present work, we will restrict ourselves to 
homogeneous systems and therefore do not consider any dependence on the position $\vec r$ in space, e.g. 
due to external potentials, in addition to the homogeneous, time-dependent electrical field that is treated as perturbation. 
The focus is on the generalization of relations which were originally derived in KT, see Sec.~ \ref{sec.2}. 
Starting from LRT, see Sec.~ \ref{sec.3},  a generalized  Boltzmann equation with a   frequency dependent  collision term is derived.
 In Sec.~\ref{sec.4},  a variational approach is applied for the solution of  the generalized linear Boltzmann equation.
Similar to the use of polynomials \cite{Chapman,Appel61,Abe71} to solve the static Boltzmann equation,
we consider moments of the single-particle distribution function to find approximate solutions. Furthermore in Sec. Sec.~ \ref{sec.2},
different limiting cases such as the dc conductivity and the high-frequency limit of the absorption coefficient are considered. 
The dynamical conductivity from KT using an energy dependent relaxation time that has often been used in the literature, is compared with the rigorous treatment
within LRT. Conclusions are drawn in Sec.~\ref{sec.6}.

\section{Kinetic Equations}
\label{sec.2}
\subsection{Single-particle distribution function}

We consider neutral Coulomb systems that consist of charged particles such as electrons and ions. 
Response to an electromagnetic field is described by the dielectric function, taken in the long-wavelength limit here, 
\begin{equation}\label{epsilon}
\lim_{k\rightarrow \infty} \epsilon(\vec k, \omega)=1+\frac{i}{\epsilon_0 \omega} \sigma( \omega)\,,
\end{equation}
or the dynamical conductivity $\sigma( \omega)$. Treating the Coulomb interaction in mean-field approximation, 
the random phase approximation 
(RPA) is obtained. To improve RPA, one 
has to include collisions. A standard way to treat collisions is the Boltzmann equation 
where the interaction between the constituents leads to the collision term. As a consequence,
a dynamical collision frequency $\nu(\omega)$ can be introduced according to a generalized Drude formula
\begin{equation}\label{sigmaom}
 \sigma( \omega)=\frac{\epsilon_0 \ompl^2}{-i\omega + \nu(\omega)}
\end{equation}
with the plasmon frequency $\ompl=\sqrt{e^2 n/(\epsilon_0 m)}$, where $n$ is the electron density and $m$ the (reduced) mass. 
The collision frequency $\nu(\omega)$ should be a complex, frequency dependent quantity in order to satisfy sum rules of the dieletctric function. This is in contrast to a static relaxation time  $\tau=1/\nu$, 
as used in the kinetic approach and will be explained in more detail below.

Taking the adiabatic approximation,   $N$ electrons interact with singly charged  
heavy ions that are considered as external potential.  The Hamiltonian with the
 electronic degrees of freedom only, is
\begin{equation}
\label{Hamiltonian}
\hat H= \sum_p E_p  \hat a_p^{\dagger}\hat a_p + \sum_{pq}V_{\rm ei}(q)  
\hat a_{p+q}^{\dagger}\hat a_p +\frac{1}{2} \sum_{p_1p_2q} V_{\rm ee}(q) 
\hat a_{p_1+q}^{\dagger} \hat a_{p_2-q}^{\dagger}\hat a_{p_2}\hat a_{p_1}
\end{equation}
with $E_p=\hbar^2p^2/(2m)$.  The interaction with the ions 
$V_{\rm ei}(\vec q)= -V(q) \sum_j^N \exp[i \vec q \cdot \vec R_j]$ 
describes Coulomb potentials $V(q)=e^2 /(\epsilon_0 \Omega_0 q^2)$ due to various ion sites $\vec R_j$ 
what leads to the structure factor $S(\vec q)= (1/N) \langle \sum_{i,j} \exp[i \vec q \cdot (\vec R_i- \vec R_j)]\rangle$. 
$\Omega_0$ is the normalization volume. The electron-electron interaction is given 
by the Coulomb interaction $V_{\rm ee}(q)= V(q)$. 
The account of the ion dynamics is straightforwardly taken into account within a two-component plasma \cite{rerrw00}, 
but  the notations become more complex and will not be given here.

For the derivation of kinetic equations,  in particular the Boltzmann equation, 
we consider the electron single-particle distribution function $f(\vec p,t)  = \tr { \hat n_p \, \hat \rho(t) }= 
\langle \hat n_p \rangle^t$
that is the quantum statistical average, taken with the non-equilibrium statistical operator  
$\hat \rho(t)$, of the single-particle occupation number operator $\hat n_p= \hat a_p^{\dagger}\hat a_p$ of momentum $\hbar \vec p$.  Considering homogeneous systems,  
the density matrix is diagonal with respect to the wave vector $\vec p$. Spin variables are not explicitly given unless it is pointed out. 
Subsequently, the single-particle distribution function does not depend on the position $\vec r$ either. 

In thermal equilibrium, the single-particle distribution function 
$ f_0(\vec p)={\rm Tr}\{ \hat n_p \,\hat \rho_0 \}$  
is calculated with the grand canconical statistical operator $\hat \rho_0=\exp[-\beta(\hat H-\mu \hat N)]/{\rm Tr}\{\exp[-\beta(\hat H-\mu \hat N)]\}$.  
Neglecting the interaction term, we find the ideal Fermi gas with distribution
$f_p = [\exp(\beta(E_p - \mu)) +1]^{-1}$.
Under the influence of an external perturbation ${\hat H}_{\rm ext}^t$, the single-particle distribution function 
$f(\vec p,t)$ is modified.
Its deviation  
\begin{equation} \label{a3}
\delta f( \vec p,t)=f(\vec p,t)-f_0( \vec p) = 
\tr{ \delta \hat n_p \,\hat \rho(t) } 
\end{equation}
from the equilibrium distribution 
$f_0( \vec p)$ is the average of the fluctuations of the single-particle occupation number 
$\delta \hat n_p = \hat n_p - f_0( \vec p)$. 
The time dependence of the single-particle distribution function $f(\vec p,t)$ 
is determined by the non-equilibrium statististical operator $\hat \rho(t)$ as shown in the following section.

Alternatively, the dynamics of the single-particle distribution function can  be determined from a 
hierarchy of equations of motions for the many-particle distribution functions. Truncating the hierarchy, 
a kinetic equation \cite{landau10} is obtained with the following structure
\begin{equation} \label{Bkineq}
\frac{\partial}{\partial t} f(\vec p,t) = {\rm D} \left[ f(\vec p,t)\right]+ {\rm C} \left[ f(\vec p,t)\right]
\end{equation}
describing drift  in the single-particle phase space via drift term ${\rm D} \left[ f(\vec p,t)\right]$,  and collisions
 that are caused by the interaction between the particles.
The collision term  $ {\rm C} \left[ f(\vec p,t)\right] $ is related to higher order 
distribution functions due to the interaction mechanisms within the system.  
To obtain closed kinetic equations, the higher distribution functions are  expressed in terms  of $f(\vec p,t) $.

In the following, we consider a  homogeneous system under the influence of an external time dependent electric field 
$ \vec E(t) $. The total 
Hamiltonian ${\hat H}_{\rm tot}^t={\hat H} +{\hat H}_{\rm ext}^t$
contains the interaction with the external field
${\hat H}_{\rm ext}^t=-e\vec E(t) \cdot \sum_i \hat{\vec r}_i$ for the electron  position operators $\hat{\vec r}_i$.  From the respective  external force $e\vec E(t)  $, 
 the drift term follows as
\begin{equation}\label{drift1}
 {\rm D} \left[ f(\vec p,t)\right]= -\frac{e}{\hbar}\vec E(t) \cdot \frac{\partial}{\partial \vec  p} \,f(\vec p,t)
\approx \frac{e \hbar}{m }\beta \,f_p (1-f_p)\,\vec   E(t)\cdot\vec p
\,
\end{equation}
in first order with respect to the external field $\vec E(t)$, with $\beta=1/( k_{\textbf{B}}T)$.
Expressions for the collision term  $ {\rm C} \left[ f(\vec p,t)\right] $ will be given below.

With the distribution function $f(\vec p,t)$, the current density is given by 
\begin{equation}
\label{jdef}
 \vec j (t) = \frac{e}{m \Omega_0} \sum_p \hbar \vec p f(\vec p,t)= \frac{e}{m \Omega_0} \vec P_1(t).
\end{equation}
The total momentum $\vec P_1(t)$ is the first moment of the distribution function. In the following, we also consider the operators of arbitrary moments 
\begin{equation}
\label{Pnu}
 \hat P_\nu = \sum_p \hbar p_E (\beta E_p)^{(\nu - 1)/2} \hat n_p
\end{equation}
where $p_E = \vec p \cdot \vec E/|\vec E|$ denotes the component of $\vec p$ in the direction of $\vec E$.

Arbitrary time dependence of an electric field can be expressed by superposition of harmonic time dependences. Within linear response, each component $\vec E(t)=\frac{1}{2} \tilde{\vec  E}(\omega)\exp(-i\omega t) + c.c.$
 causes an induced single-particle distribution function 
 \begin{equation} \label{deltaftilde}
\delta f(\vec p,t)=\frac{1}{2} \delta \tilde f(\vec p,\omega) \exp(-i\omega t) + c.c.
\end{equation}
 with  the same time dependence. The dynamical conductivity follows from $\tilde j(\omega) = \sigma(\omega) \tilde E$ as
\begin{equation} \label{cond2}
\sigma (\omega) = \frac{e }{m \tilde E}\frac{1}{ \Omega_0} \sum_p\,\hbar p_E \, \delta \tilde f(\vec p,\omega) \,.
\end{equation}
Note that all Fourier components marked with tilde, e.g. $\tilde F_p$,  
are frequency dependent in general. The dependence on $\omega$ will  be omitted  in some of the following expressions for them to be more compact. 

\subsection{Relaxation time approximation and dynamical conductivity}\label{reltime} \label{subsec_relaxtime}

To start with an analytically solvable example, we first discuss the solution of the kinetic equation (\ref{Bkineq}) for  the  Lorentz model  where the electron-electron interaction in the Hamiltonian (\ref{Hamiltonian})
is neglected. Considering  a constant electric field,  the distribution function  $f(\vec p,t)=f(\vec p)$ is static.  In the standard treatment, 
see \cite{landau10chap44},
the collision term reads
\begin{eqnarray} \label{kingl}
{\rm C_{Lorentz}} \left[ f(\vec p)\right]&=&
\sum_{p'}\left\{f(\vec p\,')w_{\rm ei}(\vec p,\vec p\,')\left[ 1-f(\vec p)\right] -f(\vec p) w_{\rm ei}(\vec p\,',\vec p) \left[ 1-f(\vec p\,') \right]\right\}.
\end{eqnarray}
The  transition rates  can be determined in Born approximation from  the golden rule,
$w_{\rm ei}(\vec p,\vec p\,')=(2\pi /\hbar) |V_{\textrm{ei}}(|\vec p-\vec p\,'|)|^2 \delta(E_p - E_{p'})$.
Since  the energy of electrons is conserved in adiabatic approximation, 
a relaxation time $\tau_p$ is introduced 
via an ansatz for the linear term of the expansion of the distribution function $f(\vec p)= f_p  -F_p \frac{1}{\beta}\frac{\partial}{\partial E_p} f_p$. In analogy to the drift term (\ref{drift1})  we assume
\begin{eqnarray} \label{relt}
\delta f( \vec p) &=&\frac{e\hbar}{m}\beta \,\tau_p \, \vec E \cdot  \vec p\,,
\end{eqnarray}
which realizes the linearity with respect to the external field $\vec E$. For isotropic systems, $\tau_p$ is a scalar depending only on the modulus of $\vec p$.
Inserting \rf{relt} into 
 the collision term (\ref{kingl}) and taking into account the detailed balance in equilibrium 
$w_{\rm ei}(\vec p,\vec p\,') f_{p'} (1-f_p)= w_{\rm ei}(\vec p\,',\vec p) f_{p} (1-f_{p'})$ as well as the energy balance of the transiton rates,  
the collision term (\ref{kingl}) is
\begin{equation}\label{crelstat}
{\rm C_{Lorentz}} \left[ f(\vec p)\right]=-\sum_{p'} w_{\rm ei}(\vec p,\vec p\,') f_p (1-f_p) (F_p-F_{p'}) =-  \delta f( \vec p)/\tau_p
\end{equation}
 For the kinetic equation (\ref{Bkineq}) with the drift term \rf{drift1} we then find
\begin{eqnarray} \label{Bkineq1}
&&\vec E \cdot \vec p = -\sum_{p'} w_{\rm ei}(\vec p,\vec p\,') \frac{ f_{p'}}{f_p}\,\vec E \cdot ( \tau_{p'} \vec p\,' -\tau_p \vec p)
=- \tau_p \sum_q w_{\rm ei}(\vec p,\vec p+\vec q) \vec E \cdot \vec q
\end{eqnarray}
with $\vec q =\vec p\,'-\vec p$.
With the golden rule for the transition rates given above and $S(q)\approx 1,\,\,|V_{\textrm{ei}}(q)|^2\approx N V^2(q) $, the energy dependent relaxation time can be calculated
 \begin{eqnarray} \label{reltime1}
  \frac{1}{\tau_p} &=&
 -\frac{2 \pi}{\hbar}\sum_{q} N V^2(q) \delta(E_p-E_{p+q})\frac{\vec E \cdot \vec q }{\vec E \cdot \vec p }.
 \end{eqnarray}
 The $\vec q$ integral in \rf{reltime1} can be performed using spherical coordinates where $\vec p$ is in $z$ direction, $\vec E$ in the $x-z$ plane.
It is convergent only in the case of a  screened Coulomb potential. Using the 
statically screened Debye potential 
\begin{eqnarray} \label{Vdeb}
V_{\rm D}(q)=\frac{e^2}{\epsilon_0 \Omega_0 (q^2+\kappa^2_{\rm D})},\qquad\qquad \kappa^2_{\rm D}=\beta n e^2 /\epsilon_0 , 
\end{eqnarray} 
we find the energy dependent collision frequency
\begin{equation} \label{coulomblog}
\nu_p=\tau_p^{-1}=n \frac{e^4}{4 \pi \epsilon_0^2} \frac{m}{\hbar^3 p^3}\left( \ln \sqrt{1+b}-\frac{1}{2} \frac{b}{1+b}\right)
\end{equation}
with $b= 4 p^2 /\kappa^2_{\rm D}$ in the Coulomb logarithm. 
The static conductivity is determined from \rf{cond2}, $\omega = 0$, as
\begin{eqnarray} \label{jel}
\sigma_{\textrm{dc,Lorentz}}&=& 
\frac{e^2 \hbar^2}{m^2}\beta \frac{1}{ \Omega_0} \sum_p \, p_E^2 \, \tau_p \,f_p(1-f_p) 
= \epsilon_0 \omega^2_{\textrm{pl}} \tau_{\rm Lorentz}=\frac{e^2 n}{m\nu_{\rm Lorentz}}\,.
\end{eqnarray}
 We introduce the average relaxation time
$\tau_{\rm Lorentz}$ and the static collision frequency $\nu_{\rm Lorentz}=1/\tau_{\rm Lorentz}$.

We are now interested in extending the static case, \rf{jel}, by evaluating the permittivity $\epsilon(\omega)$, Eq. (\ref{epsilon}), 
or the dynamical conductivity, Eq. (\ref{cond2}). 
 From the kinetic equation (\ref{Bkineq}) with the drift term \rf{drift1} we derive
the frequency dependent Boltzmann equation
\begin{eqnarray}
\label{LLpara44a}
-i\omega \delta \tilde f(\vec p,\omega)&=&   \frac{e \hbar}{m} \beta\tilde{ \vec E}(\omega) \cdot   \vec p \,f_p (1-f_p)+ {\rm C_{\rm Lorentz}} \left[ \delta \tilde f(\vec p,\omega)\right]\,.
 \end{eqnarray}

In a standard approach, see, e.g., Landau and Lifshits \cite{landau10chap44}, it is proposed to extend the static case to the dynamic case assuming that the relaxation time  is the same as in the static case, see Eq. (\ref{crelstat}).   Subsequently, the following relation is derived,
\begin{eqnarray}
&&-\left(i\omega -\frac{1}{\tau_p} \right)\delta \tilde f(\vec p,\omega) = \frac{e \hbar}{m }\beta \vec{\tilde E}(\omega)  \cdot   \vec p f_p (1-f_p)\,,
\label{LLpara44b}
\end{eqnarray}
so that for the dynamical conductivity (\ref{cond2}) follows (spin factor 2, $p_E^2\to p^2/3$ for isotropic systems)
\begin{equation}
\label{sigmaKT}
\sigma_{\rm KT}(\omega)=\frac{2 }{3 }  \frac{e^2 \hbar^2\beta}{ m^2}\int \frac{d^3 \vec p}{(2 \pi)^3} \frac{p^2 f_p(1-f_p)}{-i \omega +1/\tau_p}\,. 
\end{equation}
This result can be interpreted as a Vlassov approach where the frequency $\omega $ is replaced by a complex frequency $\omega +i/\tau_p $. 
However, the introduction of an energy dependent, static relaxation time is an approximation that cannot be applied, 
in particular, at high frequencies, where bremsstrahlung emission is expected. Note that it is not possible to give an explicit expression for a frequency dependent collision frequency as desired for a generalized Drude formula according to \rf{sigmaom}.
Furthermore, inelastic collisions such as electron-electron interactions are not taken into account by a collision time ansatz. Further evaluation of \rf{sigmaKT} is given in Appendix \ref{app4}, results are shown in Fig. \ref{fig_2b} and discussed below.

\section{Linear response equations}
\label{sec.3}
\subsection{Linear response theory}

To evaluate the response (\ref{a3}) to an external perturbation $\hat H_{\rm ext}^t$, we determine the non-equilibrium statistical operator $\rho(t)$ 
within a generalized linear response theory.
The conceptional  ideas and main expressions relevant for the further analysis of the single-particle distribution function 
will be given here according to \cite{ZMR1,Roep98,Reinholz05,Reinholz00}. 

We introduce the relevant statistical operator
\begin{equation} \label{rhorel}
\hat \rho_{\rm rel}(t)=\frac{1}{Z_{\rm rel}(t)}\e^{-\beta(\hat H-\mu \hat N) +  \sum_n F_n(t) \hat B_n}, \qquad
Z_{\rm rel}(t)= \tr {\e^{-\beta(\hat H-\mu \hat N) +  \sum_n F_n(t) \hat B_n}}\,,
\end{equation}
as a generalized Gibbs ensemble which is derived from the principle of maximum of the entropy 
\begin{equation} \label{entropy}
 S(t)=-k_B \tr{\hat \rho_{\rm rel}(t) \ln[\hat \rho_{\rm rel}(t)]},
\end{equation}
where the Lagrange parameters $\beta, \mu, F_n(t)$, which are real valued numbers, are introduced to fix the given averages
\begin{equation}
\label{selfc}
\tr{ \hat B_n \,\hat \rho(t)}=\langle \hat B_n\rangle^t =\tr { \hat B_n\,\hat \rho_{\rm rel}(t)}\,.
\end{equation}
These  self-consistency  conditions  mean that the observed averages $\langle \hat B_n\rangle^t $ are correctly reproduced 
by the hermitean $ \hat \rho_{\rm rel}(t)$. Similar relations are used in equilibrium to eliminate the Lagrange parameters $\beta$ and $\mu$. 
In linear response, the response parameters  $F_n(t)$ are considered to be small so that we can solve the
implicit relation (\ref{selfc}) expanding up to first order, 
\begin{equation} \label{rhorel2}
\hat \rho_{\rm rel}(t)=
\left[1 +  \sum_n F_n(t) \int_0^1 d\lambda \e^{-\beta \lambda(\hat H-\mu \hat N)}\delta \hat B_n 
\e^{\beta \lambda(\hat H-\mu \hat N)}\right]\hat  \rho_0\,.
\end{equation}
Note that the expansion of $ Z_{\rm rel}(t) $  in \rf{rhorel} leads to the subtraction of the equilibrium average in 
$\delta \hat B_n = \hat B_n-\langle \hat B_n\rangle_0$. 
The average fluctuations can now 
be explicitly calculated by inserting \rf{rhorel2} in \rf{selfc},
\begin{equation}
\label{selfclin}
\langle \delta \hat B_n\rangle^t=\sum_m  (\delta {\hat B}_n,\delta {\hat B}_m) F_m(t)\,,
\end{equation}
where we introduced the Kubo scalar product
\begin{equation} \label{kubokf}
( \hat A, \hat B ) = \int_0^1 {\rm d} \lambda~ \tr{\hat A \hat B^\dagger (i \hbar \beta\lambda) \hat \rho_0 }\,.
\end{equation}
The time dependence  $\hat A(t)=\e^{i\hat H t/\hbar}\hat A\e^{-i\hat H t/\hbar}$ is given by the Heisenberg 
picture with respect to the system Hamiltonian $ \hat H $, and $\hat{\dot A}=i[ \hat H, \hat A ]/\hbar $. 

A statistical operator for the non-equilibrium is constructed with the help of the relevant statistical operator (\ref{rhorel}), see App. \ref{app0}. 
Expanding up to the first order with respect to the external field $\tilde E$ and the response parameters
$\tilde F_n$, where $F_n(t)=  {\rm Re}\{\tilde F_n(\omega) \e^{-i \omega t}\}$, we arrive at the response equations
\begin{equation} \label{LBE1c}
\sum_{m} \left[\left( \hat { B}_n ;\hat{\dot B}_m\right)+\left< \hat {\dot B}_n ;\hat{\dot B}_m\right>_z - 
i \omega \left\{\left( \hat { B}_n ;\hat{ B}_m\right)+  \left<  { \hat {\dot B}_n ;\delta \hat B}_m\right>_z \right\} \right]\tilde F_{m}
=
\beta \frac{e}{m}\left\{\left( \hat { B}_n ;\hat{\vec P}\right)+  \left<  \hat {\dot B}_n ;\hat{ \vec P} \right>_z \right\}  \cdot \vec{\tilde E}\,
\end{equation}
with  $z=\omega+i \epsilon$, the total momentum of electrons $\hat{\vec P}=\sum_p\hbar \hat{\vec p} \,\hat n_p$, and the Laplace transform of the correlation functions,
\begin{equation} \label{acf}
\left< \hat A;\hat B \right>_z=\int_0^{\infty}  \textrm{d}t ~\e^{izt}\left( \hat A(t), \hat B\right)
=\int_0^{\infty}  \textrm{d}t ~\e^{izt}\int_0^1 {\rm d} \lambda~ \tr{\hat A(t-i \hbar \beta\lambda) \hat B^\dagger  \hat \rho_0 }.
\end{equation}  

Considering $N_B$ relevant observables $\hat B_n$,  \rf{LBE1c} is  a system of $N_B$ linear equations 
to determine the response parameters  $\tilde F_n$ for a given external field $\tilde E$. It is the most general 
form of LRT, allowing for arbitrary choice of relevant observables $\hat B_n$ and corresponding response parameters $F_n$.
We show below that, with respect to kinetic theory, the first two terms  on the left hand side of Eq. (\ref{LBE1c})  can be identified as a collision term, 
while the right hand side represents the drift term due to the external perturbing field.

\subsection{Generalized linear Boltzmann equations} \label{GLBE}

In kinetic theory, the non-equilibrium state is characterized by the single-particle distribution function $f(\vec p,t)$. 
In order to derive expressions in parallel to the kinetic theory, 
we choose the fluctuations $\delta \hat n_p$ of the single-particle occupation number, 
see \rf{a3}, as relevant observables $B_n$. 
The modification  of the single-particle distribution function  can then be calculated  
straight forwardly according to \rf{selfclin}
\begin{equation} \label{20}
\tr{\hat \rho_{\rm rel}(t) \,\delta \hat n_p}=\sum_{p'} \left(\delta \hat n_p, \delta \hat n_{p'}\right)  F_{p'}(t) 
 = \delta f(\vec p,t)\,.
\end{equation}
The Lagrange multipliers $F_p(t) = \tilde F_p (\omega) \exp (-i \omega t)/2  +c.c.$ are determined from the  response equations \rf{LBE1c}.
We arrive at the generalized linear Boltzmann equations
($\delta \hat{\dot{n}}_{p}= \hat{\dot{n}}_{p}$)
\begin{equation} \label{LBE}
\sum_{p'} \left[(\delta \hat n_p,\hat{ \dot{n}}_{p'})+\left< 
\hat{ \dot{n}}_{p}; \hat{ \dot{n}}_{p'} \right>_z - i \omega \left\{ (\delta \hat{n}_{p},\delta\hat n_{p'}) + 
\left<  \hat{\dot{n}}_{p}; \delta \hat{n}_{p'}\right>_z \right\}\right] \tilde F_{p'}
=
\frac{e  \hbar }{m}\beta\sum_{p''}
\left[  (\delta \hat n_p, \hat{n}_{p''}) + \left< \hat{ \dot{n}}_{p}; \hat {n}_{p''}\right>_z \right] \vec p''  \cdot \vec{\tilde E}\,.
\end{equation}
The time derivative of the position operator in $\hat H_{\rm ext}^t$  leads to the total momentum 
$\sum_i \hbar\vec p_i=m\sum_i \dot{\vec r}_i$ and subsequently to the right hand side of \rf{LBE}. 
We analyse the different terms of \rf{LBE} below and compare with the kinetic equation \rf{Bkineq}, considering the Born approximation.
Notice that this result can be extended by  introducing stochastic forces \cite{Roep98} if we go beyond the Born approximation. Further relevant observables beyond the single-particle occupation numbers can be included in order to characterize  the non-equilibrium state, such as long-living correlations and formation of bound states.  It is possible to go beyond the Boltzmann equation if higher 
correlations such as bound state formation are included into the set of relevant observables.

We give the entropy as obtained from \rf{entropy}
\begin{eqnarray}
\label{entropy1}
 S(t)&=&
-k_B \tr{\hat \rho_{\rm rel}(t) \left[- \ln[ Z_{\rm rel}(t)]-\beta (\hat H-\mu \hat N)+ \sum_p F_p(t)\,\, \hat n_p\right]}
=S_0(\beta, \mu) -
k_B \sum_p F_p(t)\,\, \delta f(\vec p,t)
\end{eqnarray} 
in first order of $F_p(t)$. The entropy in the thermodynamic equilibrium is denoted by $S_0(\beta, \mu)$. 
With Eq. (\ref{20}) we find that 
 the entropy decreases in  non-equilibrium
because $\delta S(t)=-\sum_{pp'} F_{p'}(t)  \left(\delta \hat n_{p'},\delta \hat n_p\right) F_{p}(t) \le 0$. 
The proof is given using the spectral density for $\hat F (t) = \sum_p F_p(t) \delta \hat n_p$, see  \cite{ZMR2}. With the eigenstates $(\hat H - \mu \hat N) |n \rangle = E_n |n \rangle$ of the system Hamiltonian we have
\begin{eqnarray}
\delta S(t)=-\left( \hat F(t), \hat F(t) \right)&=&
 \frac{1}{Z_0 \beta} \sum_{nm} \frac{e^{-\beta E_n}-e^{-\beta E_m}}{E_n-E_m} 
| \langle n |\hat F(t) | m \rangle |^2 \le 0\,.
\end{eqnarray}
This result corresponds to the second law of thermodynamics that the
entropy of the many-particle system exhibits
its maximum in the equilibrium state.

\subsection{Evaluation of equilibrium correlation functions, Born approximation}

Quantum statistics provide us with different methods to calculate correlation functions in thermal equilibrium 
such as perturbation theory and diagram techniques. 
Applying perturbation theory with respect to the interaction, Wick's theorem can be used. 
We find in lowest order for  the Kubo scalar product, \rf{kubokf},
\begin{eqnarray}
\left({\hat n}_{p} ,\hat n_{p'}\right )&=& \tro{\hat a_{p'}^{\dagger}\hat a_{p'} \hat a_p^{\dagger}\hat a_p}= 
f_{p'}f_{p}+  f_p(1-f_p)\delta_{p p'}
\end{eqnarray}
so that $\left(\delta \hat n_p,\delta \hat n_{p'}\right)=\left( \delta \hat n_p,\hat n_{p'}\right)=  f_p(1-f_p)\delta_{p p'}$.
The remaining Kubo scalar product vanishes, $(\delta \hat n_p,\hat{\dot{n}}_{p'})=0$,  as shown from the Kubo identity \rf{kuboidentity}
with $\hat C =\delta \hat n_p$, and $\langle [n_{p'},n_p] \rangle_0=0 $ after cyclic invariance of the trace.

For the deviation of the single-particle occupation numbers from equilibrium we find from \rf{20}
that $\label{20b}
 \delta f( \vec p,t)= F_p(t)  f_p(1-f_p) $
which is equivalent to the expansion (\ref{relt}) in kinetic theory. Thus, we solved the self-consistency condition (\ref{selfc}) to eliminate the Lagrange parameters $F_p(t)$. According to (\ref{deltaftilde}), the Fourier components 
\begin{equation} 
\label{Fpfp}
 \delta \tilde f( \vec p,\omega)=  f_p(1-f_p) \tilde F_p(\omega)
\end{equation}
 are complex amplitudes, containing in general a phase factor.

The equation of motion that leads to the generalized linear Boltzmann equation (\ref{LBE})
allows to relate the response to the external field. 
The right-hand side is the drift term that contains the external field. In Born approximation, 
we can neglect the correlation function $ \left<\hat{ \dot{n}}_{p}; \hat {n}_{p''}\right>_z  $
because it is of higher order of interaction compared with $(\delta \hat n_p,\hat{n}_{p''})$. Then, the right-hand side of  \rf{LBE} 
reads 
\begin{equation}
\label{Dp} 
 {\rm D}_p= \frac{e \hbar}{m } \beta f_p (1-f_p)\, \vec  p \cdot \tilde{\vec E} 
\end{equation}
in agreement with \rf{drift1}.
By the same argument we have the term due to the explicit time dependence
\begin{equation} \label{frequterm}
-\sum_{p'} i \omega \left[ (\delta\hat n_p,\delta \hat{n}_{p'}) + 
\left< \hat{\dot{n}}_{p}; \delta \hat{n}_{p'}\right>_z \right]\tilde F_{p'} =-  i \omega  \delta \tilde f( \vec p,\omega)=-i \Omega_p \tilde F_{p}\,,
\end{equation}
with $\Omega_p=\omega f_p(1-f_p)$. Note that the correlation function $ \left< \hat{ \dot{n}}_{p};\hat {n}_{p''}\right>_z  $ 
is eliminated introducing stochastic forces \cite{Roep98,ZMR2} so that the result $- i \omega  \delta \tilde f( \vec p,\omega)$ 
holds also beyond the Born approximation.

The remaining term in  \rf{LBE} describes the collision integral,
\begin{equation} \label{collterm}
{\rm C}_p=-\sum_{p'}   \left< 
\hat{ \dot{n}}_{p};\hat{ \dot{n}}_{p'} \right>_{\omega+i \epsilon} \tilde F_{p'} =-\sum_{p'} {\cal L}_{pp'}(\omega) \tilde F_{p'} .
\end{equation}
It is evaluated in Born approximation, see Appendix \ref{app1}, 
\begin{details}
 \begin{equation}
 {\rm St}_p[\delta \tilde f ({\vec p})]=\sum_{p'}  \tilde F_{p'} \left< \hat{ \dot{n}}_{p'};
 \hat{ \dot{n}}_{p} \right>_z =\sum_{p'} P_{pp'}  \left[ \frac{1}{f_{p'}(1-f_{p'})}\delta \tilde f ({\vec p'})-  \frac{1}{f_{p}(1-f_{p})}\delta \tilde f ({\vec p})\right] \,,
 \end{equation}
\end{details} 
with the generalized Onsager coefficients ${\cal L}_{pp'}(\omega)={\cal L}_{pp'}^{\rm ei}(\omega)+{\cal L}_{pp'}^{\rm ee}(\omega)$, leading to
\begin{details}
 \begin{eqnarray}
 {\cal L}_{pp'}^{\rm ei}(\omega) & = & - \frac{ 1}{\hbar^2}   |V_{\rm ei}(\vec p'-\vec p)|^2  \frac{ f_{p'}-f_p}{E_{p'}-E_p} 
 \left[ \pi \delta(\omega+(E_p-E_{p'})/\hbar)+\pi \delta(\omega-(E_p-E_{p'})/\hbar)\right.\nonumber\\&& \left.
 +i \frac{\cal P}{\omega+(E_p-E_{p'})/\hbar}+i \frac{\cal P}{\omega-(E_p-E_{p'})/\hbar} \right].
 \end{eqnarray}
\end{details} 
\begin{eqnarray}
\label{Lei}
{\cal L}_{pp'}^{\rm ei}(\omega) & = & - \frac{ 1}{\hbar^2} \sum_q  |V_{\rm ei}(q)|^2  \frac{ f_{p}-f_{p+q}}{\beta (E_{p+q}-E_p)} 
\left[ \pi \delta\left(\omega+\frac{1}{\hbar }(E_p-E_{p+q})\right)+\pi \delta\left(\omega-\frac{1}{\hbar }(E_p-E_{p+q})\right)\right.\nonumber\\&& \left.
+i \frac{\cal P}{\omega+(E_p-E_{p+q})/\hbar}+i \frac{\cal P}{\omega-(E_p-E_{p+q})/\hbar} \right] 
\left[\delta_{p',p+q}-\delta_{p',p} \right]\,, \label{LLei}
\end{eqnarray}
\begin{eqnarray}
\label{Lee}
{\cal L}_{pp'}^{\rm ee}(\omega) & = & - \frac{ 1}{\hbar^2} \sum_{p_1,q}  |V_{\rm ee}(q)|^2  
\frac{f_p f_{p_1} (1-f_{p_1-q}-f_{p+q})- f_{p+q}f_{p_1-q}(1-f_{p_1}-f_p)}{\beta(E_{p+q}+E_{p_1-q}-E_{p_1}-E_p)} \nonumber\\&& 
\left[ 
 \frac{i}{\omega+i \epsilon +\Delta_{p,p_1,q}}+\frac{i}{\omega+i \epsilon-\Delta_{p,p_1,q}} \right] 
 \left[\delta_{p',p+q}+\delta_{p',p_1-q}-\delta_{p',p_1}-\delta_{p',p} \right],  \label{LLee}
\end{eqnarray}
where $\Delta_{p,p_1,q}=  (E_{p+q}+E_{p_1-q}-E_{p_1}-E_p)/\hbar$. Exchange contributions have been discarded, see Appendix \ref{app1}. The decomposition of ${\cal L}_{pp'}^{\rm ee}(\omega)$ in real and imaginary part is 
analoguous to ${\cal L}_{pp'}^{\rm ei}(\omega)$.

In conclusion, the generalized linerized Boltzmann equation (\ref{LBE}) can be given in the same way as assumed in the relaxation time approach, see \rf{LLpara44a},
\begin{equation}
\label{kinBG}
- i \omega \delta \tilde f ({\vec p},\omega)= \frac{e \hbar}{m } \beta f_p (1-f_p)\, \vec  p \cdot \tilde{\vec E}- \sum_{p'} {\cal L}_{pp'}(\omega) \tilde F_{p'}={\rm D}_p +{\rm C}_p[\delta \tilde f ({\vec p},\omega)]
 \end{equation}
with \rf{frequterm} and  the drift term (\ref{Dp}),  after  replacing the response parameters $\tilde F_p$ in the collision term (\ref{collterm}) by the single-particle distribution according to \rf{Fpfp}. This holds  for arbitrary frequencies $\omega$ and degeneracy, see Appendix A. 
 At zero frequency, the collision integral (\ref{kingl}) of the Lorentz plasma  is recovered if calculations are taken  in Born approximation and restricted to the electron-ion interaction only. 
At arbitrary frequencies, the collision integral becomes a complex quantity in contrast to the scalar relaxation time. 
Real  and imaginary part are connected via Kramers-Kronig relations.
The Born approximation can be improved in a systematic way if the correlation functions are 
evaluated in higher orders with respect to the interaction. A Kubo-Greenwood formula can be 
derived that expresses the collision term by T matrices \cite{Roep98,Reinholz00}.

\section{Solution of the generalized linear Boltzmann equation}\label{sec.4}

\subsection{ Variational principle}

Having derived explicit expression for the Onsager coefficients ${\cal L}_{pp'}$ in Born approximation, Eqs. (\ref{Lei}), (\ref{Lee}), we can now determine the response parameters  by solving the generalized linear Boltzmann equation  (\ref{kinBG}) given as  
\begin{equation}
\label{kinBGF}
- i \Omega_p  \tilde F_p(\omega) ={\rm D}_p- \sum_{p'} {\cal L}_{pp'}(\omega) \tilde F_{p'}(\omega).
\end{equation}
 As a further constraint on the response parameters  $\tilde F_p$, we consider the entropy leading to a variational problem as follows.

We determine
 the time derivative of the entropy, \rf{entropy1}. The time dependent term reads
\begin{eqnarray}
\label{entrchange}
&& \frac{d}{dt}S(t)=-2 \sum_p \frac{1}{ f_p(1-f_p)} \delta f( \vec p,t) \delta \dot f( \vec p,t) 
=-\frac{1}{2} \sum_p \frac{1}{ f_p(1-f_p)} [\delta \tilde f( \vec p) \e^{-i \omega t}+c.c.] 
[- i \omega \delta \tilde f( \vec p) \e^{-i \omega t}+c.c.]\nonumber
\\ && 
=-\frac{1}{2} \sum_p \left[ \tilde F_p \e^{-i \omega t}+ \tilde F^*_p \e^{i \omega t}\right] 
\left[{\rm D}_p[\tilde {\vec E}]
\left( \e^{-i \omega t}+ \e^{i \omega t}\right)-
  \sum_{p'} {\cal L}_{pp'}(\omega) \tilde F_{p'} \e^{-i \omega t}-
  \sum_{p'} {\cal L}^*_{pp'}(\omega) \tilde F^*_{p'} \e^{i \omega t}\right]
\end{eqnarray}
if we insert the Boltzmann equation (\ref{kinBG}) for $-i \omega \delta \tilde f ({\vec p})$ for the last line. 
Oscillating terms $\propto \e^{2i\omega t},  \e^{-2i\omega t}$ arise that disappear in the time average. The remaining terms cancel, which can be directly seen, if replacing $\delta \tilde f( \vec p) $ by the Langrange multipliers $\tilde F_p$ using \rf{Fpfp}. 
Thus the total entropy is constant in the average over a period of time, $d \bar S(t)/dt = 0$.
However, even  in the time average, there is an entropy production which is dissipated as entropy export due to the external field in the drift term. We have
\begin{equation}
\label{bilanz}
\frac{d \bar S(t)}{dt}=\dot S_{\rm ext}+\dot S_{\rm int} 
= -\frac{e \hbar}{2m } \beta\sum_p  \tilde F^*_p  f_p (1-f_p)\, \vec  p \cdot \tilde{\vec E}  +
\frac{1}{2}\sum_{pp'} \tilde F^*_p   {\cal L}_{pp'}(\omega) \tilde F_{p'} +c.c.
 =0
\end{equation}
Therefore, let us consider the functional
\begin{equation} \label{funcS}
\dot S_{\rm int}[\tilde G_p] = \sum_{pp'} \tilde G^*_p {\cal L}_{pp'}(\omega) \tilde G_{p'} +c.c.
\end{equation}
for any function $\tilde G_p$ that obeys the constraint
\begin{equation}
\label{auxcon}
 \sum_p \tilde G^*_p \left[-{\rm D}_p- i \Omega_p  \tilde G_p+\sum_{p'}  {\cal L}_{pp'}(\omega) \tilde G_{p'}\right] =0
\end{equation}
that can be considered as an integral over the Boltzmann equation (\ref{kinBGF}). 
It is easily shown that  the time averaged change of entropy \rf{entrchange} vanishes for arbitrary functions $\tilde G_p$ that obey the constraint (\ref{auxcon}).
 The  maximum of the functional $\dot S_{\rm int}[\tilde G_p]$ occurs at $  \tilde G_p = \tilde F_p$ which is the solution of the linear Boltzmann equation (\ref{kinBGF}),  see App. \ref{app2} for the proof. 

This is a generalization of the Kohler variational principle  \cite{kohler1,kohler2} for arbitrary frequencies $\omega$.
It can be related  to the principle of extremum of  entropy production given by Prigogine and Glansdorff \cite{prigogine}.
The static case $\omega=0$ has been considered in Refs. \cite{Appel61,kohler1,kohler2,ChrisRoep85}. Some attempts to extent this to arbitrary frequencies can be found in \cite{SamHoj71}, but, to our knowledge,  a consistent approach has  not been given until now.

In order to apply the variational principle given here, one can consider a class of trial functions  $\tilde G^{(N_\nu)}(\Phi_\nu;\vec p)=\sum_{\nu=1}^{N_\nu} \Phi_\nu g_\nu(\vec p)$ with respect to an arbitrary but finite ($N_\nu$) set of linear independent  functions $ g_\nu(\vec p)$. 
Determining the extremum of $\dot S_{\rm int}[ \Phi_\nu]$  leads to an optimal set of parameters $\Phi_\nu^{\rm opt}=F_\nu^{(N_\nu)}$. The extension of the class of trial functions to an infinite number of functions then gives  the exact result $\tilde F_p=\lim_{N_\nu \rightarrow \infty}\sum_{\nu=1}^{N_\nu} F_\nu^{(N_\nu)} g_\nu(\vec p)$.

Alternatively, the  relevant observables $\hat n_p$ are replaced by  a reduced set of $N_\nu$ relevant observables   $\hat B_\nu=\sum_p g_\nu(\vec p)\hat n_p$. The solution of the finite system  of linear equations (\ref{LBE1a}) then gives the Lagrange multipliers $F_\nu$, that can be expressed in terms of determinants. This leads to identical results as  for the variational principle. In previous papers we used a finite number of moments $g_\nu(\vec p)=\hbar  p_E  (\beta E_p)^{(\nu-1)/2}$ according to the general moments (\ref{Pnu}). An alternative basis set would be the Sonine polynomials \cite{Chapman} that are appropriate in the static, nondegenerate limit. It has been shown that, within perturbation expansion \cite{Reinholz89,Redmer97}, results are converging with an increasing number  of moments used. 

\subsection{One-moment Born approximation}

In lowest approximation, we choose with  $\tilde G_p=F_1 g_1(p) = F_1 \hbar p_E $ the first moment of the distribution function (\ref{Pnu}) as  trial function.
The variational parameter $F_1$ is fixed by the auxiliary condition (\ref{auxcon}) where we insert \rf{Dp} and $\Omega_p$ from  \rf{frequterm}, and we find
\begin{eqnarray}
\label{dynkinequc}
&&\sum_p F_1 \hbar p_E  \frac{e \hbar }{m} \beta f_p (1-f_p) p_E \tilde E=
-i\omega \sum_p (F_1 \hbar p_E)^2 f_{p} (1-f_{p})  
-\sum_{p,p'} F_1 \hbar   p_E {\cal L}_{pp'}^{\rm ei}(\omega)  F_1 \hbar   p_E' \,.
\end{eqnarray}
The electron-electron collisions do not contribute in the one-moment approach because of conservation of total momentum.
We  assume the  general structure of the variational parameter
\begin{equation}
\label{onemoment}
F_1= \frac{e \beta }{  m } \frac{1}{[ -i \omega+  \nu_{\rm D}(\omega)]  } \tilde E\,.
\end{equation}
After some calculations given in Appendix \ref{app3},  we find  the collision frequency for the case of the statically screened Coulomb potential  \rf{Vdeb},  and $S(q) \approx 1$,
\begin{equation}\label{nudegen}
\nu_{\rm D}(\omega)=i g_{\rm degen} \int_0^\infty dy \frac{y^3}{(y^2+\bar n)^2} \int_{-\infty}^\infty \frac{dx}{x} 
\frac{1}{w+i \varepsilon -x } 
\ln \left[\frac{1+e^{-(x/y-y)^2+\beta \mu}}{1+e^{-(x/y+y)^2+\beta \mu}}\right]\,
\end{equation}
with 
\begin{equation}
g_{\rm degen}=\frac{1}{48 \pi^{4}}\frac{e^4m}{ \epsilon_0^2 \hbar^3},\qquad  
w= \frac{\beta \hbar \omega}{4 }, \qquad \bar n= \frac{\beta \hbar^2 \kappa^2_{\rm D}}{8 m }\,,
\end{equation}
which is valid for any degeneracy. 
In the non-degenerate limit  $\beta \mu \ll 1$, we can expand the logarithm.
With $e^{\beta \mu}= n (2 \pi \beta \hbar^2/m )^{3/2}/2 = n \Lambda^3/2 $ and spin factor 2, 
we find
\begin{equation} \label{nuclassic}
\nu_{\rm D}(\omega)=i g \,n \int_0^\infty dy \frac{y^4 }{(y^2+\bar n)^2} \int_{-\infty}^\infty dx  \frac{1-e^{-4 xy} }{xy (w-xy +i \varepsilon) } e^{-(x-y)^2} 
\end{equation}
 with $g= \Lambda^{3}\,g_{\rm degen}/2$.  

The dynamical conductivity (\ref{cond2}) can now be calculated with \rf{Fpfp} and the optimized Lagrange parameter \rf{onemoment} so that  $\tilde F_p= F_1 \hbar p_E $. We find
\begin{equation}
\label{drudesum}
\sigma_{\rm D}({\omega})=\frac{e }{m \tilde E} F_1 \frac{1}{ \Omega_0} \sum_p\, (\hbar p_E )^2 \,  f_p(1-f_p).
\end{equation}
For isotropic systems, the sum is evaluated as $\sum_p\, (\hbar p_E )^2 \,  f_p(1-f_p) = N m/ \beta$, see  App. \ref{app3}. Inserting  the derived expression (\ref{onemoment}) 
 we obtain  a generalized 
Drude type expression, \rf{sigmaom}, 
\begin{equation}\label{Drude1}
 \sigma_{\rm D}( \omega)=\frac{\epsilon_0 \ompl^2}{-i\omega + \nu_{\rm D}(\omega)}
\end{equation}
 for the dynamical conductivity. 
The comparison with $\sigma_{\rm KT}$, \rf{sigmaKT}, will be performed in the following Section.

It is instructive to investigate 
 the alternative approach where  only moments of the distribution function $\hat P_\nu$, \rf{Pnu}, are taken as relevant observables $\hat B_n$, instead of the fluctuations $\delta \hat n_p$ of the single-particle occupation operator as originally introduced in Subsec. \ref{GLBE}.
Taking the component of the  total momentum of the electrons  $\hat P_1=\sum_p \hbar p_E   \hat n_p$ in the direction of $\vec E$
as an one-moment approach, we have with \rf{jdef}, (\ref{selfclin})
\begin{equation}
 \tilde j = \frac{e}{m \Omega_0} \langle \hat{P}_{1}\rangle F_1=\frac{e}{m \Omega_0} (\hat{P}_{1}, \hat P_1) F_1\,.
\end{equation}
The generalized linear Boltzmann equation  (\ref{LBE}) is now reduced to a single equation 
that reads in Born approximation ($\langle \hat P_1 \rangle_0=0$ in thermal equilibrium)
\begin{equation} \label{P0}
\left[ \left< \hat{ \dot{P}}_{1};\hat{ \dot{P}}_{1} \right>_{\omega+ i \epsilon} - i \omega  ( \hat{P}_{1},\hat P_1)  \right]  F_{1} =
 (\hat{P}_{1}, \hat P_1) \frac{e  }{m} \beta \tilde E\,
\end{equation}
containing  force-force correlation functions as the collison term. With $(\hat{P}_{1}, \hat P_1)= N m/\beta$, see App. \ref{app3} and  the statically screened interaction \rf{Vdeb}, the expressions for the dynamical conductivity, \rf{Drude1}, and the corresponding dynamical collision frequency  
\begin{equation}
 \nu^{(P_1)}_{\rm D}(\omega)= \frac{\beta}{m\,n\, \Omega_0} \langle\hat {\dot P}_1;\hat {\dot P}_1 \rangle_{\omega+ i \epsilon}
\end{equation}
is obtained that coincides
 with the results \rf{nudegen} and  \rf{nuclassic} given above.
 This is a preliminary result of the LRT based on the one-moment Born approximation. Going beyond the Born approximation, we denote  $ \nu^{(P_1)}(\omega)= \beta/(m\,N) \langle\hat {\dot P}_1;\hat {\dot P}_1 \rangle_{\omega+ i \epsilon} $ as collision frequency of the one-moment approach. 
Systematic treatments of the perturbation expansions are performed with the help of Green's function techniques. 
In particular, the Gould-DeWitt approximation for $\nu^{(P_1)}(\omega)$ has been performed that accounts for the correction of long-range interaction 
by dynamical screening and considers strong collisions at short ranges \cite{Reinholz00,rerrw00}.

\subsection{Higher moment approaches}\label{subsx}

An improvement of the dynamical conductivity (\ref{Drude1}) can be achieved by extending the set of trial functions 
or  relevant observables within the variational approach  or the relevant statistical operator, respectively. Using  higher order moments  $\hat P_\nu$, \rf{Pnu},
of the distribution function, converging expressions 
are obtained for the transport coefficients \cite{Redmer97,Roepke89}. In particular, higher moments are needed in order to take into account electron-electron collisions. Taking higher order moments into account, the change of the dynamical conductivity can be represented 
by a complex function $r(\omega)$  so that $\nu(\omega) = r(\omega)\nu^{(P_1)}(\omega)$ \cite{rerrw00,Reinholz05,rr00},
\begin{equation}\label{Drude2}
 \sigma( \omega)=\frac{\epsilon_0 \ompl^2}{-i\omega + r(\omega)\nu^{(P_1)}(\omega)}\,.
\end{equation}

As a special case, we discuss the two-moment approach with $\hat P_1,\,\hat P_3$ as relevant observables (i.e. particle current and energy current). The account of these two functions in $p$ space allows for a better variational approach to the single-particle distribution function. For the electrical current density we have with \rf{jdef}, (\ref{selfclin})
\begin{equation}\label{2current}
\tilde j = \frac{e}{m \Omega_0} \langle \tilde P_1 \rangle = \frac{e}{m \Omega_0} \left\{ (\hat{P}_{1}, \hat P_1)  F_1 +  (\hat{P}_{1}, \hat P_3)  F_3 \right\} = \sigma( \omega) \tilde E\,.
\end{equation}
According to the response equations (\ref{LBE1c}), see also \rf{P0}, the Lagrange parameters $F_1, F_2$ are determined via the generalized linear Boltzmann equations, taken in  Born approximation, 
\begin{eqnarray} \label{P2}
&&\left[ \left< \hat{ \dot{P}}_{1};\hat{ \dot{P}}_{1} \right>_{\omega+ i \epsilon} - i \omega  ( \hat{P}_{1},\hat P_1)  \right]  F_{1}
+  \left[ \left< \hat{ \dot{P}}_{3};\hat{ \dot{P}}_{3} \right>_{\omega+ i \epsilon} - i \omega  ( \hat{P}_{1},\hat P_3)  \right]  F_{3} =
 (\hat{P}_{1}, \hat P_1) \frac{e  }{m} \beta \tilde E\,\nonumber \\
 &&\left[ \left< \hat{ \dot{P}}_{3};\hat{ \dot{P}}_{1} \right>_{\omega+ i \epsilon} - i \omega  ( \hat{P}_{3},\hat P_1)  \right]  F_{1}
+ \left[ \left< \hat{ \dot{P}}_{3};\hat{ \dot{P}}_{3} \right>_{\omega+ i \epsilon} - i \omega  ( \hat{P}_{3},\hat P_3)  \right]  F_{3}=
(\hat{P}_{3}, \hat P_1) \frac{e}{m} \beta \tilde E \,.
\end{eqnarray}

As shown in Appendix \ref{app3}, we have $( \hat{P}_{1},\hat P_1)=Nm/\beta, \,( \hat{P}_{1},\hat P_3)=( \hat{P}_{3},\hat P_1)= \frac{5}{2} Nm/\beta, \,( \hat{P}_{3},\hat P_3)=  \frac{5}{2} \frac{7}{2}Nm/\beta\,.$
Using Cramers rule, the response parameters $F_1, F_2$ are expressed in terms of the electrical field $\tilde E$ and correlation functions.
For the dynamical conductivity, \rf{2current}, we find after algebraic manipulations  the expression \rf{Drude2} with
\begin{equation}\label{2sigma}
r( \omega) =  \frac{\frac{5}{2}i \omega N \frac{m}{\beta}- \left< \hat{ \dot{P}}_{3};\hat{ \dot{P}}_{3} \right>_{\omega+ i \epsilon} 
+\frac{ \left< \hat{ \dot{P}}_{1};\hat{ \dot{P}}_{3} \right>_{\omega+ i \epsilon}  \left< \hat{ \dot{P}}_{3};\hat{ \dot{P}}_{1} \right>_{\omega+ i \epsilon}}{  \left< \hat{ \dot{P}}_{1};\hat{ \dot{P}}_{1} \right>_{\omega+ i \epsilon}}}{\frac{5}{2}i \omega N \frac{m}{\beta}-\frac{25}{4}\left< \hat{ \dot{P}}_{1};\hat{ \dot{P}}_{1} \right>_{\omega+ i \epsilon}+\frac{5}{2}\left< \hat{ \dot{P}}_{1};\hat{ \dot{P}}_{3} \right>_{\omega+ i \epsilon}+\frac{5}{2}\left< \hat{ \dot{P}}_{3};\hat{ \dot{P}}_{1} \right>_{\omega+ i \epsilon}-\left< \hat{ \dot{P}}_{3};\hat{ \dot{P}}_{3} \right>_{\omega+ i \epsilon} }.
\end{equation}
Evaluation of the correlation functions occurring in the renormalization factor  $r(\omega)$ in Born approximation is given in Appendix \ref{app4}.

\vspace*{0.4cm}
\begin{figure}[htp] 
\begin{center}
 \includegraphics[width=6.8cm]{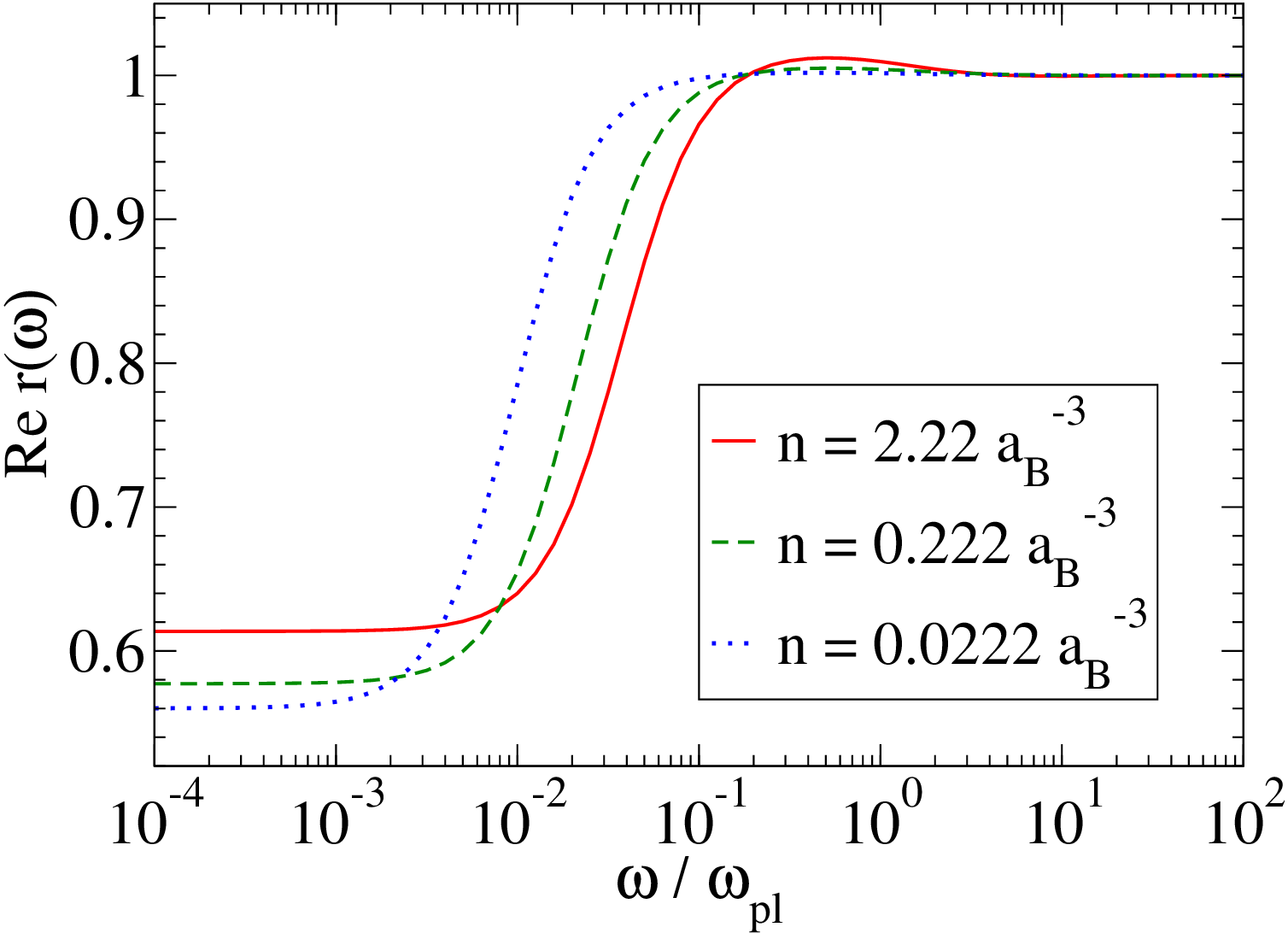}
(a)
\hspace{0.1cm}
 \includegraphics[width=7cm]{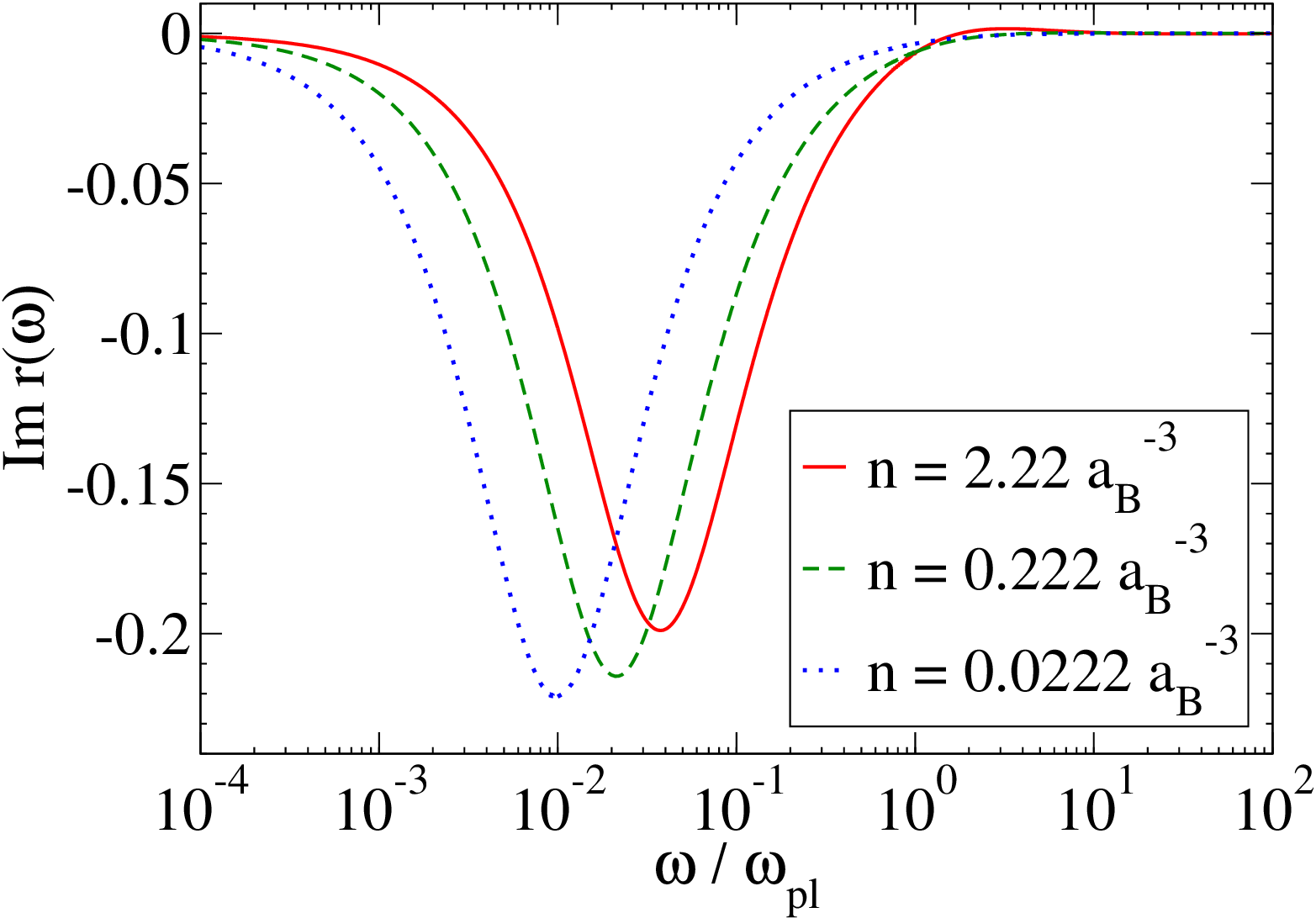}
 (b)
\end{center}
  \caption{ Frequency dependence of (a) the real part  and (b)  the  imaginary part  
of the renormalization factor \rf{2sigma}. 
Hydrogen plasmas at temperature  $T = 42.13$ Ryd = 573 eV (solar core) and three different electron densities $n$ are considered.}
\label{fig_1b}
\end{figure}

Results for the renormalization factor at solar core conditions  and  lower densities are shown in Fig. \ref{fig_1b}. At solar core conditions ($T=573$ eV = 42.13 Ryd, $n = 1.51 \times 10^{25}\, {\rm cm}^{-3} = 2.22\, a_B^{-3}$), we have a weakly interacting (plasma parameter $\Gamma= e^2/(4 \pi \epsilon_0 k_BT) \,(4 \pi n/3)^{1/3}=0.1$) and  nearly degenerate
(degeneration parameter $\Theta= 2 m k_BT/\hbar^2 (3 \pi^2n)^{-2/3}=1.3$)  plasma. At the lower densities, the plasma becomes more classical. 
At high frequencies (i.e. large compared with the inverse relaxation time), $r (\omega)$ approaches 1, and higher moments of the momentum distribution 
that describe the deformation from a shifted Fermi distribution are not relevant. 
In the static case, the real part Re $r(0)$ shows the effect of $e-e$ collisions according to the Spitzer result \cite{Reinholz89,Reinholz05}. 
Since the Coulomb logarithm (\ref{coulomblog})  depends on the density, in addition to the correct prefactor also the density dependence of the Coulomb logarithm occurring in the different moments is 
seen. Only in the very low-density limit the different Coulomb logarithms cancel.

So far we evaluated the equilibrium correlation functions occurring in the generalized linear Boltzmann equation \rf{LBE} 
with the help of perturbation theory. Thus we solved a kinetic equation using a variational approach or a reduced set of relevant observables.  
Note that one can go beyond the kinetic equation that treats the single-particle distribution function 
by considering fluctuations in the two-particle states  as additional relevant observables in the generalized LRT \cite{Roep88,EssRoep98}. 

\subsection{ Limiting cases  }
\subsubsection{Zero-frequency limit: Static conductivity} %

We rewrite the dynamical collision frequency (\ref{nudegen}) in a symmetric form by transforming $x \to -x$ in half of the expression and using the Dirac identity, 
\begin{eqnarray}
\label{nudegen1}
\nu_{\rm D}(\omega)
&=&
\frac{g_{\rm degen}}{2}  \int_0^\infty dy \frac{y^3}{(y^2+\bar n)^2} \int_{-\infty}^\infty \frac{dx}{x} 
\left\{\pi \delta(x-w)+\pi \delta(x+w)-i\frac{{\cal P}}{x-w} +i\frac{{\cal P}}{x+w} \right\}
\ln \left[\frac{1+e^{-(x/y-y)^2+\beta \mu}}{1+e^{-(x/y+y)^2+\beta \mu}}\right]\nonumber \\ &&{}
\end{eqnarray}
The pricipal values compensate in the static case $w=0$. 
After expanding for small $x$, $e^{-(x/y-y)^2+\beta \mu}\approx e^{-y^2+\beta \mu}[1+2x]$, 
the integral over $x$ can be performed with the result
\begin{eqnarray}
\label{nudegen2}
\lim_{\omega\rightarrow 0} \, \nu_{\rm D}(\omega)&=&
2 \pi g_{\rm degen} \int_0^\infty dy \frac{y^3}{(y^2+\bar n)^2} \frac{1}{e^{y^2-\beta \mu}+1}\,.
\end{eqnarray}
Note that only $e-i$ collisions contribute to the one-moment Born approximation.

First we discuss the Lorentz model. It is solved for the static case in KT using an energy dependent relaxation time. The dc conductivity in Born approximation for the one-moment approach (\ref{Drude1}), $\sigma_{\rm D}(0)= 
\epsilon_0 \ompl^2/ \nu_{\rm D}(0)$,  is not identical with $\sigma_{\rm dc}$ obtained from \rf{jel} with the Coulomb logarithm (\ref{coulomblog}), because $1/ \nu_{\rm D}(0) \ne \tau_{\rm Lorentz}$. This difference stems from the fact that in the one moment approach with the variational parameter $F_1$ the $p$ dependence is specified as $g_1(p)= \hbar p_E$. 
The  $p$ dependence necessary for the Lorentz model to reproduce   the result for the relaxation time approach is given by $g_4(p)$, see \rf{Pnu}, and is only roughly approximated by $g_1(p)$ within the interval of relevance. However, If we add further moments $g_\nu(p)$, not necessarily including $g_4(p)$, the approximation of the exact $p$ dependence is improving. This has already been extensively investigated, see Refs. \cite{Redmer97,Adams07}. The dc conductivity within LRT follows from Eq. (\ref{Drude2}) as 
$\sigma( 0)=\epsilon_0 \ompl^2/[ r(0)\nu^{(P_1)}(0)]$ with the static renormalization factor $r(0)$. 
The collision frequency $\nu^{(P_1)}(0)$ improves the Born approximation $\nu_{\rm D}(0)$ if further effects like dynamical screening and strong collisions are included.

The equivalence of the KT and LRT for the Lorentz plasma in the static case $\omega =0$ can be shown rigorously by inspection of the kinetic equation. Taking the linearized Boltzmann equation (\ref{kinBG}) with the collision term (\ref{collterm}), (\ref{Lei}) in the static limit, 
\begin{eqnarray}
-\frac{e\hbar }{m} \, \beta f_p(1-f_p) \vec p \cdot \vec {\tilde E} 
&=& -\sum_{p'} {\cal L}_{pp'}^{\rm ei}(\omega) F_{p'}
\\ \label{statickinequ} 
 &=&
\frac{2\pi }{\hbar}~\sum_{q}~|V_{\rm ei}(q)|^2~\delta\left(E_{p+q}-E_{p}\right)\frac{f_p-f_{p+q}}{\beta(E_{p+q}-E_p)}
  \left[ F_{p+q}-F_p\right]\nonumber \\
 &=&
\frac{2\pi }{\hbar}~\sum_{q}~|V_{\rm ei}(q)|^2~\delta\left(E_{p+q}-E_{p}\right)
  \left[ \delta \tilde f(\vec p+\vec q) - \delta \tilde f(\vec p)\right]\,
\end{eqnarray}
where the expression (\ref{Fpfp}) is used to insert the change of the single-particle distribution function $\delta \tilde f(\vec p)$ after expanding $f_{p+q}- f_p \approx  (\partial/\partial \beta E_p) f_p = - \beta (E_{p+q}-E_p) f_p (1-f_p)$. 
This equation coincides with the equation of motion for the single-particle distribution function (\ref{kingl}), that is
obtained in the static case from KT and is solved using the relaxation time ansatz. 

Considering the electron-ion plasma, it should be pointed out that the relaxation time approximation is not applicable if electron-electron collisions are relevant. 
In contrast, $\sigma (0)$ obtained from LRT contains also the contribution of electron-electron collisions as given by \rf{Lee} in the static limit. For this, the static renormalization factor $r(0)$ can be evaluated from \rf{2sigma}. In particular, it gives the correct Spitzer result if strong collisions are included \cite{Roep88,Redmer97,Adams07}, see also Sec. \ref{subsx}.

\subsubsection{High-frequency limit: inverse bremsstrahlung absorption}

The dielectric function $\epsilon (\omega)=[n_r(\omega)+i c/(2 \omega) \alpha (\omega)]^{1/2}$ determines the refraction index $n_r(\omega)$ as well as the absorption coefficient $\alpha(\omega)$. We consider the long-wavelength limit where the transversal and longitudinal dielectric function coincide. The dielectric function or the optical conductivity $\sigma (\omega)$ can be used to calculate the inverse bremsstrahlung absorption. In the high-frequency limit, where  $n_r(\omega) \approx 1  $ and $\omega \gg \nu $, we have  
\begin{equation}
 \alpha(\omega) = \frac{\omega}{c \,n_r(\omega)} {\rm Im}\, \epsilon(\omega) \approx \frac{\omega_{\rm pl}^2}{\omega^2 c} {\rm Re}\, \nu(\omega)
\end{equation}
so that the inverse bremsstrahlung absorption coefficient is directly related to the dynamical collision frequency obtained above from the solution of the Boltzmann equation.

Bremsstrahlung radiation is described by the Bethe-Heitler expression resulting from QED in second order of interaction \cite{IZ,Bekefi}.
In the non-relativistic limit and for soft photons, the absorption coefficient for a hydrogen plasma ($Z_i=1$) is given by \cite{Fortmann06,FRRW}
\begin{equation}
 \alpha^{\rm Born}(\omega) = \frac{64 \pi^{3/2}n^2 \sqrt{\beta} }{3 \sqrt{2} m^{3/2} \hbar c\, \omega^3} \left( \frac{e^2}{4 \pi \epsilon_0}\right)^3
\sinh\left(\frac{1}{2}\beta\hbar \omega\right) K_0\left(\frac{1}{2}\beta\hbar \omega\right)
\end{equation}
where $K_0(x)=\int_0^\infty dt \exp[-x \cosh( t)] = \int_0^\infty dy\,\exp[-y^2-x^2/(4y^2)]/y  $ is the modified Bessel function of zeroth order.

Generalized LRT gives the same result. We use  the collision frequency \rf{nuclassic} in the nondegenerate case. 
 At finite frequencies $\omega$, the integral with $\bar n =0$ is no longer divergent at $y=0$. Therefore, the screening of the Coulomb potential  can be neglected ($\bar n=0$).
We find \cite{Reinholz05,Fortmann06} 
\begin{equation}
 \alpha^{\rm Born}(\omega) = \frac{16\sqrt{2} \,\pi^{7/2} n^2 \sqrt{\beta}}{(3 m)^{3/2}  \hbar c\, \omega^3} \left( \frac{e^2}{4 \pi \epsilon_0}\right)^3
\left(1-e^{-\beta \hbar \omega} \right) g^{\rm Born}_{ff}(\omega)\,,
\end{equation}
with the free-free Gaunt factor  in Born approximation 
\begin{equation}
 g^{\rm Born}_{ff}(\omega)=\frac{\sqrt{3}}{\pi^2} e^{\beta \hbar \omega/2}K_0\left(\frac{1}{2}\beta\hbar \omega \right).
\end{equation}
The well-known Kramers formula for the inverse bremsstrahlung absorption \cite{Kramers23} results with the Gaunt factor $g^{\rm Kramers}_{ff}(\omega)=1$.

The one-moment Born approximation can be improved taking into account dynamical screening, 
strong collisions, and higher moments of the distribution function, as discussed earlier. However, in the high-frequency limit, the dynamical screening is not of relevance. The frequency dependence of the renormalization factor has been discussed in \cite{Reinholz05}, see also Fig. \ref{fig_1b},  and converges to 1 in the high-frequency limit.
Strong collisions have been considered and lead to the famous Sommerfeld result for the Gaunt factor \cite{Som49,Wier01}.
For dense plasmas, the account of ion correlation $S(\vec q)$ [see Eq. (\ref{Hamiltonian})] has a major effect and can directly included in the Born approximation \cite{Totsuji85}.

The standard treatment of the kinetic equation using a relaxation time ansatz, see Subsec. \ref{subsec_relaxtime}, fails to describe inverse bremsstrahlung absorption. The frequently used expression (\ref{sigmaKT}) for the dynamical conductivity, or the corresponding expression for the dielectric function, are restricted to the low-frequency region since a static, but energy dependent relaxation time cannot be applied to the high-frequency region. Different approaches using Fermi's golden rule have been used \cite{landau10} to derive expressions for the emission of radiation.
A common treatment unifying both limiting cases, $\omega \to 0$ and $\omega \to \infty$, is missing in KT within the relaxation time approximation.

In contrast,  our approach within LRT covers the entire frequency regime consistently. Note that it can also be applied to the degenerate case and to the relativistic regime, see \cite{Hoell03}. An important feature of the LRT is the possibility to include medium effects in dense plasmas such as the Landau-Pomeranchuk-Migdal effect  \cite{Fortm07}.

\subsection{Dimensionless dynamical conductivity}
\label{subsx1}

In the following we use 
Rydberg units where  $\hbar = 1,\, a_B=1,\, m=1/2,\, e^2/(4 \pi \epsilon_0)=2,\,k_B=1$. The temperature $T$  is then given 
 in Ryd =13.6 eV and the electron density $n$ in $a_B^{-3}$.
We introduce dimensionless quantities 
$ \omega^*=\omega/\omega_{\rm pl}\equiv w\, T/ \sqrt{\pi n}$ and 
\begin{equation}
\label{Rydb}
\sigma^*(\omega)=\frac{e^2 \beta^{3/2} m^{1/2}}{(4 \pi \epsilon_0)^2} \; \sigma(\omega) \,.
\end{equation}

\begin{figure}[htp] 
\begin{center}
 \includegraphics[width=7.5cm]{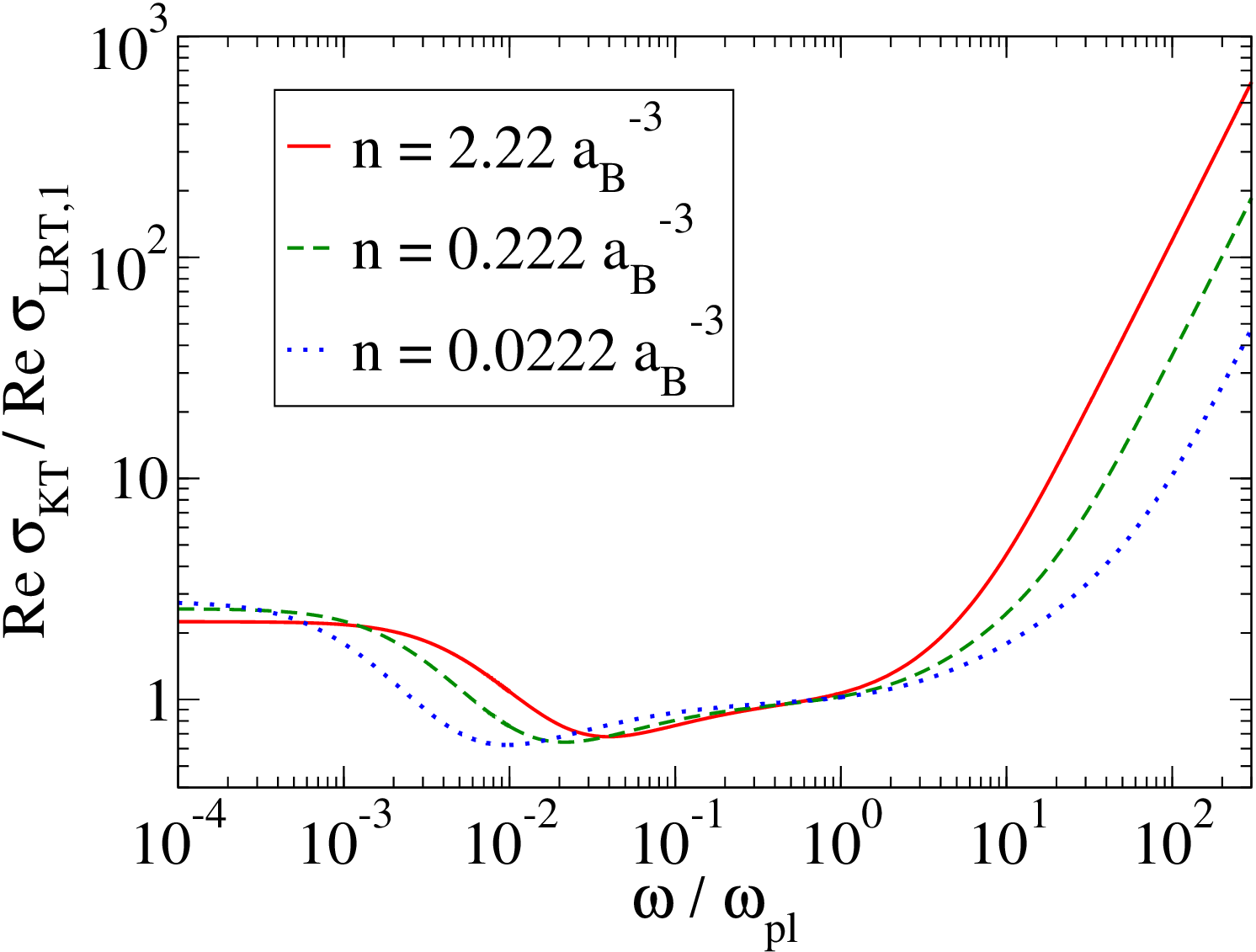}\,\,\,\,
(a)
\hspace{0.1cm}
 \includegraphics[width=7.5cm]{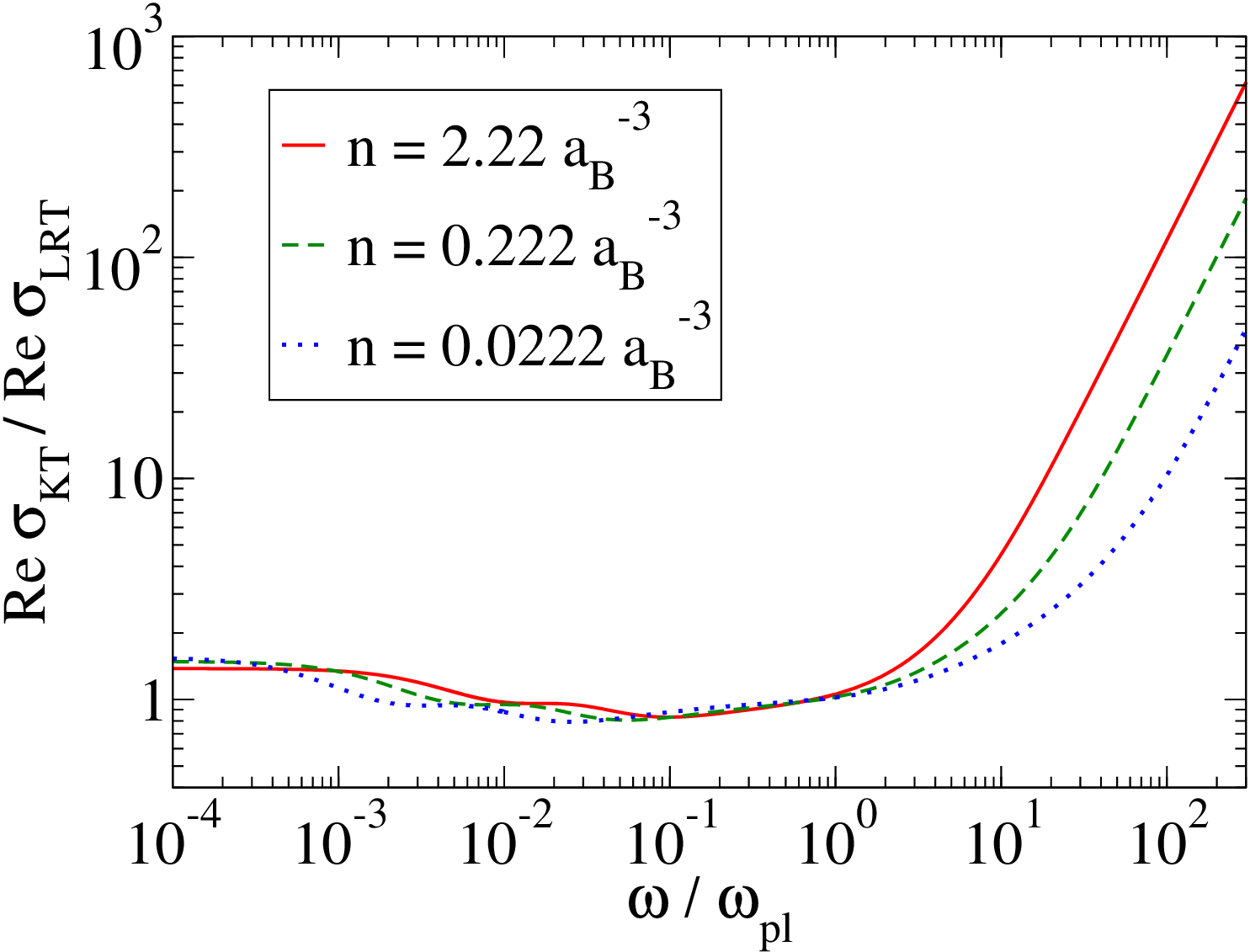}
 (b)
\end{center}
  \caption{  Ratio of the real part of the dynamical conductivity, calculated within relaxation time ansatz \rf{sigmaKT*} in comparison to 
generalized linear Boltzmann equation (LRT) in (a)  one-moment approximation \rf{sigmaLRT*1} and (b) two-moment approximation \ref{sigmaLRT*}. Hydrogen plasmas at temperature  $T = 42.13$ Ryd = 573 eV (solar core) and three different electron densities $n$ are considered.}
\label{fig_2b}
\end{figure}

In Fig. \ref{fig_2b}(a), the ratio of the kinetic theory to the linear response theory is shown for the real part of the dynamical conductivity 
at various parameter values. The one-moment approximation is used, corresponding to the force-force correlation function. 
In Fig. \ref{fig_2b}(b), the renormalization factor is included.
In the low-frequency limit, deviations are shown that are due to the inclusion of e-e contributions. 
We give the limits   of the  expressions \rf{sigmaKT*} and \rf{sigmaLRT*1}, given in Appendix \ref{app4}, in the static case
\begin{eqnarray}
\sigma^*_{\rm KT}(\omega = 0)&=&\frac{2^{5/2}}{\pi^{3/2}}\frac{1}{\Lambda_{\rm KT}}\nonumber\\
\sigma^*_{\rm LRT,1}(\omega = 0)&=&\frac{3}{2^{5/2}\pi^{1/2}}\frac{1}{\Lambda_{\rm LRT,1}}\,.
\end{eqnarray}
In both approaches, the Coulomb logarithm  behaves like $\lim_{n \to 0}\Lambda \sim -\frac{1}{2} \ln n$ in the low-density limit. 
At finite densities, different expressions are observed.
The prefactor of the inverse Coulomb logarithm takes the value 1.015 for the Lorentz model that corresponds to the KT in relaxation time approximation. The Spitzer value 0.591 is approached in the LRT considering the Born approximation (0.2992 in the one-moment case,  
0.5781 in the two-moment case). This quick convergence is known from the literature, see \cite{Redmer97}.
The inclusion of the third moment of the momentum distribution takes electron-electron interaction as well as transport of heat
into account.

In the high frequency limit, we find from  \rf{sigmaKT*} and \rf{sigmaLRT*1}  the asymptotic expansions
\begin{eqnarray}
\textrm{Re}\,\sigma^*_{\rm KT}(\omega \to \infty)&=&\frac{16\sqrt{ 2}n}{3\sqrt{\pi}T^3}\Lambda_{\rm KT}\frac{1}{\omega^2}\nonumber\\
\textrm{Re}\, \sigma^*_{\rm LRT,1}(\omega \to \infty)&=&\frac{\sqrt{ 2} n^{1/4}}{3 \pi^{5/4} T^{3/2}}\frac{1}{\omega^{7/2}}\,.
\end{eqnarray}
The ratio between KT and LRT behaves as $\omega^{3/2}$. Thus, in the high frequency limit, the  ratio diverges, see Fig. \ref{fig_2b}. 
In conclusion, above the plasma frequency the kinetic approach becomes essentially wrong.

\vspace*{0.5cm}

\section{Conclusion}\label{sec.6}

Considering the interaction of radiation with matter, often a dielectric function or dynamical conductivity is used that is derived
from kinetic theory using an energy dependent relaxation time, see \rf{sigmaKT} and \rf{sigmaKT*}. However, this expression is valid only
for elastic collisions of electrons so that electron-electron collisions cannot be included. Furthermore, the frequency dependence is not correctly
described. In particular, the high-frequency behavior has a wrong dependence on $\omega$ and fails to describe inverse bremsstrahlung. We developed an alternative approach that is free from these shortcomings. 

We have derived a generalized linear Boltzmann equation \rf{LBE} that is valid for any frequencies and at arbitrary degeneracy. 
Besides electron-ion interaction, also electron-electron interaction is included. The drift term and the collision term 
are expressed in terms of equilibrium correlation functions that are, in general, complex quantities. In order to apply this  approach consistently, one has to deal with two problems as follows. 

Firstly, the correlation functions can be evaluated numerically or, using quantum statistical methods, in perturbation theory. 
As simplest approximation, we considered the Born approximation, see \rf{Drude1} with \rf{nudegen}, \rf{nuclassic}, and \rf{sigmaLRT*}. 
This leads to analytic expressions that are tractable to be used for simple evaluations.

Secondly,  solving the generalized linear Boltzmann equation, a variational principle has been applied that optimizes the single-particle distribution function within a subspace of trial functions. In particular, we considered a finite number of moments of the distribution function.
The single-moment treatment gives  a result for the dynamical conductivity that is improved if higher moments
of the distribution functions are taken into account. The contribution of higher moments is represented by the renormalization
factor $r(\omega)$ that is, in general, a complex quantity. The high-frequency limit is not modified 
by the inclusion of higher moments and reproduces the well-known results for bremsstrahlung. 
The static limit converges to the Spitzer result for the conductivity with the inclusion of higher 
moments that desribe also the contribution of electron-electron interaction.

We compared both approaches for different plasma properties. In the case of the Lorentz plasma that takes into account only elastic scattering 
of electrons by the ions, the correct static conductivity is obtained in KT using an energy dependent relaxation time. 
To get this result in LRT, the variational solution with only the lowest moment $P_1$ is not sufficient, and higher moments should be considered. In particular, the inclusion of the fourth moment $P_4$ alone gives  the exact result for the static conductivity. 
the solution of KT with an energy dependent relaxation time becomes increasingly inappropriate with higher frequencies. In contrast, the expressions obtained from LRT are applicable at any frequency.

Considering the more realistic case of the electron-ion plasma, the relaxation time ansatz to solve the kinetic equation breaks down.
The inclusion of electron-electron collisions where the single-particle energy is not conserved represents no problem in LRT. The exact results for the transport coefficients in the low-density limit given by the Spitzer formula are reproduced by the LRT, in contrast to KT. The correct treatment of inverse bremsstrahlung shows that LRT is valid in the entire frequency domain, in contrast to KT using the energy dependent relaxation time that cannot reproduce the correct frequency dependence of the optical conductivity. 

Starting from a general LRT, a linearized Boltzmann kinetic equation has been obtained, and the relation to the results of th the relaxation time approach in the KT discussed. We restricted ourselves to a two-moment Born approximation. Possible improvements as pointed out throughout the paper are summarized here again as an outlook to further considerations and calculations.
 
\begin{itemize}
\item
Taking the single-particle occupation number $ n_p$ as relevant observables $B_n$, the deviations 
from equilibrium  $\langle \hat n_p \rangle^t-f_0(\vec p)$ describe the non-equilibrium state. 
 The set of relevant observables can be extended by including initial state correlations,
in particular the formation of 
bound states. This is straight forward in a general version of the LRT, see e.g. \cite{Adams07,Rrn95b}
Sophisticated approaches have been worked out to show conservation of total 
energy and the systematic inclusion of correlations and bound state formation, 
using non-equilibrium Green's function theory \cite{Kremp,Mor01} or  within
generalized linear response theory \cite{Roep88,Mor99}.
This is of relevance to investigate  partially ionized plasmas, but also allows for the treatment of quasiparticle formation and the Debye-Onsager relaxation effect.

\item
In linear response theory, the drift term and the collision term are expressed in terms of equilibrium correlation functions. 
They can be evaluated numerically or within perturbation theory, if we expand with respect to the interaction.
The Born approximation is improved if higher orders with respect to the interaction are taken into account. 
The technique of thermodynamic Green´s functions has been used for the evaluation of equilibrium correlation functions \cite{rerrw00,Reinholz05}.
The binary collision approximation is obtained 
if ladder diagrams are summed up. 
Dynamical screening results from the summation of ring diagrams. Perturbation expansions are more efficient 
if correlations are already included in the set of relevant observables so that they dan't have to be generated  by a dynamical treatment, i.e.,
by considering higher order perturbation expansions. As example, we refer to the formation of bound states discussed above. 
Instead of finding their influence using higher orders of perturbation theory, we can treat them as new degrees of freedom introducing
 the corresponding relevant observables, e.g. their disribution function or a finite number of moments. Then, memory effects become less important, 
and the Markov approximation can be used, e.g., introducing stochastic forces \cite{ZMR2}.

\end{itemize}

Equilibrium correlation functions that determine the transport coefficients can be calculated for arbitrary frequencies, 
degeneracy, electron-electron collisions, and including collective excitations. 
The frequency dependence and further aspects are disregarded if a relaxation time is introduced. 
The relaxation time approach  is exact only in the case of elastic scattering, for instance of electrons by ions in the adiabatic limit. 
Electron-electron scattering as well as finite frequencies of the electric field  cannot be treated by the relaxation time ansatz. 
Thus, the generalized linear Boltzmann equation obtained from linear response theory reproduces some well-known benchmarks 
such as the Spitzer result for the static conductivity of the fully ionized plasma or the Kramers formula for the bremsstrahlung.

\appendix

\section{Derivation of the response equations}
\label{app0}

The hermitean observables $ \hat B_n$ are assumed to conserve the 
total particle number so that the entropy operator $\hat H-\mu \hat N$ is replaced by the system's  Hamiltonian  $\hat H$ in the $\lambda$ dependence of the relevant statistical operator (\ref{rhorel2}). 
Note that the averages  are 
calculated with the equilibrium statistical operator that is known to us, and quantum statistical methods can be 
applied such as Green function techniques or numerical simulations to evaluate it. 
Thus, in linear response theory the Lagrange multipliers $F_n(t)$ can 
be eliminated using equilibrium correlation fuctions. 

The relevant statistical operator serves as initial condition to determine the non-equilibrium statistical operator 
$\rho(t)$.  Further correlations are build up by the dynamical evolution \cite{ZMR2} with the total Hamiltonian 
$\hat H_{\rm tot}^t=\hat H+\hat H_{\rm ext}^t$,
\begin{equation} \label{statop}
\hat \rho(t)= \lim_{\epsilon \to 0} \epsilon \int_{-\infty}^t dt' \e^{-\epsilon (t-t')} \hat U(t,t')
\hat \rho_{\rm rel}(t') \hat U^\dagger(t,t') 
\end{equation}
with the time evolution operator $\hat U(t,t')$ given by $ i \hbar (\partial/\partial t)\hat U(t,t')=
\hat H_{\rm tot}^t \hat U(t,t')$ and $\hat U(t,t)=1  $. 
The external perturbation to the system's Hamiltonian $\hat H$ shall have the general form 
$\hat H_{\rm ext}^t=\sum_j h_j(t) \hat A_j $. Decomposition of the time dependence of the field into Fourier components 
$h_j(t)=\tilde h_j(\omega) e^{-i \omega t}/2+c.c.= {\rm Re} \{\tilde h_j(\omega) e^{-i \omega t}\}$ is particularly convenient in linear response 
since the reaction of the system is the superposition of the reaction to  different spectral components 
of the external perturbation. Subsequently,   the time dependence of the response to each component will 
have the same frequency in the stationary case, i.e. $F_n(t)=  {\rm Re}\{\tilde F_n(\omega) \e^{-i \omega t}\}$. In the following, we consider a fixed 
value $\omega$ for the frequency of the external perturbation.

We now perform a partial integration of the statistical operator (\ref{statop}) and 
linearize with respect to the external fields $\tilde h_j$ and the response parameters $\tilde  F_n$, 
\begin{eqnarray} \label{statopa}
\hat \rho_{\rm irrel}(t)&=&\hat \rho(t)-\hat \rho_{\rm rel}(t)=- \lim_{\epsilon \to 0}  \int_{-\infty}^t dt' \e^{-\epsilon (t-t')} 
\e^{-i\hat H (t-t')/\hbar}\left\{ \frac{i}{\hbar}\left[ \hat H_{\rm ext}^{t'},\hat \rho_0\right]\right. \nonumber\\&& \left. +
\sum_n  \int_0^1 d\lambda \e^{-\beta \lambda(\hat H-\mu \hat N)}\left( \frac{i}{\hbar}\left[ \hat H, \delta \hat B_n \right]\,F_n(t')
+ \delta \hat B_n\,\frac{\partial}{\partial t'}F_n(t')\, \right)\e^{\beta \lambda(\hat H-\mu \hat N)} \right\} 
\e^{i\hat H (t-t')/\hbar} \hat \rho_0\,.
\end{eqnarray}
According to Eq. (\ref{selfc}) we have ${\rm Tr} \left\{ B_n\, \hat \rho_{\rm irrel}(t) \right\} = 0$,
for details see \cite{Reinholz05,Roep98}.
Finally, applying the Kubo identity
\begin{equation}  \label{kuboidentity}
\beta \int_0^1 d\lambda  e^{-\lambda \beta \hat H} [\hat C, \hat H]  e^{\lambda \beta \hat H} \hat \rho_0 = 
\int_0^1 d\lambda  \frac{d}{d \lambda} \hat C (- i \hbar \beta \lambda) \hat \rho_0 = [ \hat C , \hat \rho_0]\,,
\end{equation}
with 
$\hat C=\hat H_{\rm ext}^{t'}$,
\begin{details}
\begin{equation}
 \frac{i}{\hbar}\left[ \hat H_{\rm ext}^{t'},\hat \rho_0\right]=
\beta  \sum_{j}  \int_0^1 {\rm d} \lambda~  \hat {\dot A}_j (i \hbar \beta\lambda)   h_j(t') \hat \rho_0\,,
\end{equation}
\end{details}
we find an expression that relates the response parameters $\tilde F_n$ 
to the external fields $\tilde h_j$,
\begin{equation} \label{LBE1a}
\sum_{m} \left[\left< \hat { B}_n ;\hat{\dot B}_m\right>_z - 
i \omega  \left<  { \hat { B}_n ;\delta \hat B}_m\right>_z \right]\tilde F_{m}
=
- \beta \sum_{j} \left<  \hat { B}_n ;\hat{ \dot A}_j \right>_z  \tilde h_j  ,
\end{equation}
where the Laplace transform of the correlation functions (\ref{acf}) has been introduced.
After partial integration, $-i z \left< \hat A;\hat B \right>_z=  (\hat{ A},\hat B)-\left< \hat A;\hat {\dot B} \right>_z
= (\hat{ A},\hat B)+\left< \hat {\dot A};\hat  B \right>_z$, we arrive at the response equations (\ref{LBE1c}) with the external perturbation
${\hat H}_{\rm ext}^t=-e \hat{\vec R}  \cdot  \vec E(t), \quad  \hat{\vec R}= \sum_i \hat{\vec r}_i,$ and $  \hat{\dot{\vec R}}=\hat {\vec P}/m$.

\section{Evaluation of the collision term}
\label{app1}
We evaluate the Onsager coefficient ${\cal L}_{pp'}(\omega)=\left<  \dot{n}_{p'}; \dot{n}_{p} \right>_{\omega +i\epsilon}$  which  occurs in the collision term \rf{collterm} of 
the linearized equation of motion for the single-particle distribution function \rf{LBE}. 
Inserting the time derivative of the occupation number  (using $V^*(-q)=V(q)$), 
\begin{eqnarray}
\hat{ \dot{n}}_{p}&=&\frac{i}{\hbar}  [\hat H,{ \hat n}_p] = \frac{i}{\hbar}\sum_q V_{\rm ei}(q)
\left[\hat a_{p+q}^{\dagger}\hat a_p-\hat a_p^{\dagger}\hat a_{p+q}\right]\nonumber\\&&
+ \frac{i}{\hbar}\sum_{p'q} V_{\rm ee}(q)\left[\hat a_{p+q}^{\dagger}\hat a_{p'-q}^{\dagger}\hat a_{p'}\hat a_p-
\hat a_p^{\dagger}\hat a_{p'}^{\dagger}\hat a_{p'-q}\hat a_{p+q}\right] \label{ndot}
\end{eqnarray}
 into  Eqs.\,(\ref{kubokf}), (\ref{acf}), we evaluate  the correlation functions for the electron-ion contribution in Born approximation
\begin{eqnarray} \label{b3}
\left< \hat{ \dot{n}}_{p};\hat{ \dot{n}}_{p'} \right>^{\rm ei}_{\omega+i\epsilon}=&&
-\frac{1}{\hbar^2}~\sum_{q q'}~V_{\rm ei}(q)V_{\rm ei}(q')~\int^{\infty}_0  \textrm{d}t ~\e^{i(\omega+i\epsilon) t} \int_0^1 {\rm d} \lambda~
\\  \nonumber
&& \times \left\{ 
\left[ \tro{\hat a_{p+q}^{\dagger} \hat a_{p} \hat a_{p'+q'}^{\dagger} \hat a_{p'}} - \tro{\hat a_{p+q}^{\dagger} \hat a_{p} \hat a_{p'}^{\dagger}  
\hat a_{p'+q'}}\right]
\e^{\frac{i}{\hbar}\left(E_{p+q}-E_{p}\right)(t-i \hbar \beta \lambda)}
\right.
\\  \nonumber
&& \left.- \left[ \tro{\hat a_{p}^{\dagger} \hat a_{p+q} \hat a_{p'+q'}^{\dagger} \hat a_{p'}} -\tro{\hat a_{p}^{\dagger} \hat a_{p+q} 
\hat a_{p'}^{\dagger} \hat a_{p'+q'}}\right]
\e^{\frac{i}{\hbar}\left(E_{p}-E_{p+q}\right)(t-i \hbar \beta \lambda)}
\right\}\,.
\end{eqnarray}
The $\lambda$ intergral can be executed. The application of the Wick theorem to the quantum statistical averages $\tro{\dots}$  
leads to $\delta$ functions, in particular $q=-q'$. Contributions with $q=0$ cancel. We assume isotropic interaction  $V(\vec q)=V(-\vec q)$ and obtain
\begin{eqnarray}
{\cal L}_{pp'}^{\rm ei}(\omega)&=&
-\frac{1 }{\hbar^2}~\sum_{q}~|V_{\rm ei}(q)|^2 \frac{e^{\beta(E_{p+q}-E_p)}-1}{\beta( E_{p+q}-E_p)}f_{p+q}(1-f_p)
 \frac{-1}{i( \omega+i\epsilon)+i(E_{p+q}-E_p)/\hbar} \left[ \delta_{p', p+q}- \delta_{p', p}\right] \nonumber \\
&&+\frac{1 }{\hbar^2}~\sum_{q}~|V_{\rm ei}(q)|^2 \frac{e^{\beta(E_{p}-E_{p+q})}-1}{\beta( E_{p}-E_{p+q})}f_{p}(1-f_{p+q}) \frac{-1}{i( \omega+i\epsilon)+i(E_{p}-E_{p+q})/\hbar} \left[ \delta_{p', p}- \delta_{p', p+q}\right] \,\nonumber \\
&=&-\frac{1 }{\hbar^2}~\sum_{q}~|V_{\rm ei}(q)|^2 \frac{f_p-f_{p+q}}{\beta( E_{p+q}-E_p)}\nonumber \\
&&\times \left\{ \frac{i}{ \omega+i\epsilon+(E_{p+q}-E_p)/\hbar}+\frac{i}{ \omega+i\epsilon-(E_{p+q}-E_p)/\hbar} \right\}
 \left[ \delta_{p', p+q}- \delta_{p', p}\right]\,,   \label{npnp}
\end{eqnarray}
using $ (e^{\beta (E_{p'}-E_p)}-1)f_{p'}(1-f_p)=f_p-f_{p'}$. Subsequently, the Onsager coefficient can be given as \rf{LLei}.

With this result, the collision term \rf{collterm} for the Lorentz plasma reads
\begin{eqnarray}
\label{dynkinequb1}
{\rm C}_p^{\rm ei}=
  \frac{1}{\hbar^2} \sum_{q}  |V_{\rm ei}(q)|^2 \frac{f_p-f_{p+q}}{\beta (E_{p+q}-E_p)}  
\left\{ \frac{i}{\omega+i\epsilon+(E_{p}-E_{p-q})/\hbar}+
 \frac{i}{\omega+i\epsilon-
(E_{p}-E_{p-q})/\hbar} \right\}(\tilde F_{ p+q}-  \tilde F_{ p})
\end{eqnarray}
which is now a frequency dependent and complex quantitiy. We can eliminate the Lagrange multiplier $\tilde F_p$ according \rf{Fpfp} in order  to express the collision integral in terms of the single-particle distribution function. 

A similar calculation gives the electron-electron contribution in Born approximation
\begin{eqnarray}
&&{\cal L}_{pp'}^{\rm ee}(\omega)=
-\frac{ 1 }{\hbar^2}~\sum_{p_1,q}~V_{\rm ee}(q) V_{\rm ee, ex}(q;p,p_1)\nonumber \\
&&\left\{\frac{e^{\beta (E_{p+q}+E_{p_1-q}-E_{p_1}-E_p)}-1}{\beta  (E_{p+q}+E_{p_1-q}-E_{p_1}-E_p)} \frac{i}{ \omega+i\epsilon- 
(E_{p+q}+E_{p_1-q}-E_{p_1}-E_p)/\hbar}f_{p+q}f_{p_1-q}(1-f_{p_1})(1-f_{p})  
 \right.
 \nonumber\\
&&\left. +\frac{e^{\beta (E_{p}+E_{p_1}-E_{p_1-q}-E_{p+q})}-1}{\beta  (E_{p}+E_{p_1}-E_{p_1-q}-E_{p+q})} \frac{i}{ \omega+i\epsilon-  
(E_{p}+E_{p_1}-E_{p_1-q}-E_{p+q})/\hbar}f_{p}f_{p_1} (1-f_{p_1-q})  (1-f_{p+q})  \right\}
\nonumber \\&&\times\left[ \delta_{p', p+q}+ \delta_{p', p_1-q}- \delta_{p', p_1}- \delta_{p', p}\right]\,,  \label{npnpel}
\end{eqnarray}
where $V_{\rm ee, ex}(q;p,p_1)
= V_{\rm ee}(q)-\delta_{\sigma_1,\sigma_2}V_{\rm ee}(|\vec p_1-\vec p-\vec q|)$ is the 
exchange interaction with $\sigma_i$ denoting the spin explicitely. The respective Onsager coefficient can be given as \rf{LLee}. It is easily seen from the final expressions (\ref{LLei}), (\ref{LLee}) that the real part  of the Onsager coefficient  ${\cal L}_{pp'}(\omega)= {\cal L}_{pp'}^{\rm ei}(\omega)+{\cal L}_{pp'}^{\rm ee}(\omega)$  is  non-negative, ${\rm Re}\, {\cal L}_{pp'}(\omega)\ge 0$. 

\section{Proof of the variational solution}
\label{app2}

To begin with,  we show that  the entropy production \rf{funcS},  
\begin{equation} \label{B1}
 \dot S_{\rm int}[\tilde G_p] =  \sum_{pp'} \tilde G^*_{p}  ({\cal L}_{pp'}(\omega)+{\cal L}^*_{p'p}(\omega))\tilde G_{p'} 
= \sum_{pp'}  \tilde G^*_{p} \left< \hat{ \dot{n}}_{p};\hat{ \dot{n}}_{p'} \right>_{\omega +i\epsilon}\tilde G_{p'} 
\end{equation}
as a functional of an arbitrary $\tilde G_p$  is positive definite. Using the spectral density of the operator 
$\hat G=\sum_p \tilde G_p \hat {\dot n}_p$, we find

\begin{equation}\label{B2}
 \dot S_{\rm int}[\tilde G_p] = \left<  \hat G ;\hat G \right>_{\omega +i\epsilon}=\frac{1}{Z_0 } \sum_{nm}\frac{e^{-\beta E_m}-e^{-\beta E_n}}{\beta (E_n-E_m) }
 \pi \delta \left(\omega +\frac{1}{\hbar }(E_n-E_m)\right) |\langle n |\hat G | m \rangle|^2 \ge 0\,. 
\end{equation}

 Now we consider the functional \rf{B1} for the function $(\tilde G_p-\tilde F_p)$ and decompose
 \begin{eqnarray}
\label{sdot}
\dot S_{\rm int}[(\tilde G_p-\tilde F_p)]&=&\dot S_{\rm int}[\tilde G_p] -\sum_{pp'}[ \tilde G^*_p  {\cal L}_{pp'}(\omega)
\tilde F_{p'} +c.c.]
-\sum_{pp'}[ \tilde F^*_p   {\cal L}_{pp'}(\omega) \tilde G_{p'} +c.c.] +\dot S_{\rm int}[\tilde F_p]
\end{eqnarray}
Making use of the  constraint \rf{auxcon}, the first  contribution is expressed as
\begin{equation} \label{term1}
\dot S_{\rm int}[\tilde G_p] =  \sum_p\left [ \tilde G^*_p+ \tilde G_p \right]{\rm D}_p\,,
\end{equation}
the terms with $ i \Omega_p$ compensate. Since $\tilde F_{p}$ solves the linear Boltzmann equation (\ref{kinBGF}), the 
second contribution is transformed into
\begin{equation}\label{term2}
\sum_{pp'} \tilde G^*_p {\cal L}_{pp'}(\omega) 
\tilde F_{p'}+c.c.= \sum_p \tilde G^*_p {\rm D}_p+\sum_p i \Omega_p \tilde G^*_p\tilde F_{p}+c.c.
\end{equation}
For the transformation of the third term, we use the symmetry  $ {\cal L}_{pp'}(\omega)= {\cal L}_{p'p}(\omega) $ due to detailed balance which can be seen easily from the explicit expressions \rf{npnp} and \rf{npnpel}. Furthermore, the proof of  the reciprocitivity condition ${\cal L}_{pp'}(\omega)={\cal L}^*_{pp'}(-\omega)$ can be shown generally using   the eigenstates 
 $ |n \rangle$ 
of the system Hamiltonian,
\begin{eqnarray}
 {\cal L}_{pp'}(\omega)&=&\frac{1}{\hbar^2} \frac{1}{Z_0 \beta} \sum_{nm} \frac{e^{-\beta E_n}-e^{-\beta E_m}}{E_n-E_m} 
\frac{(E_n-E_m)^2}{i \omega - \epsilon -(i/\hbar)(E_n-E_m) } \langle n |\hat n_p | m \rangle  \langle m |\hat n_{p'} | n \rangle\,
\end{eqnarray}
interchanging $n$ and $m$. Finally, we find
\begin{eqnarray} \label{term3}
&&
\sum_{pp'} \tilde F^*_p  {\cal L}_{pp'}(\omega) \tilde G_{p'} = \sum_{pp'}[ \tilde G^*_{p'} {\cal L}^*_{pp'}(\omega)\tilde F_p]^*
=
\sum_{pp'}[ \tilde G^*_{p}  {\cal L}_{pp'}(-\omega) \tilde F_{p'}]^*= \sum_p  [{\rm D}_p+i\Omega_p \tilde F^*_{p}]\tilde G_{p}\,.
\end{eqnarray}
We sum up all contributions in  \rf{sdot} using the Eqs. (\ref{term1}), (\ref{term2}), (\ref{term3}),
\begin{equation}
 \dot S_{\rm int}[(\tilde G_p-\tilde F_p)]=\dot S_{\rm int}[\tilde F_p] - \dot S_{\rm int}[\tilde G_p] \ge 0\,.
\end{equation}
This is a positive definite expression due to \rf{B2}. Thus we find that the entropy production is maximal  if the trial function 
$\tilde G_p$ is the solution $\tilde F_p$ of the Boltzmann equation. 

\section{Evaluation of Eq.~(\ref{dynkinequc})}\label{app3}

We execute the $\vec p$ integration on the left hand side of \rf{dynkinequc} with  $p_E^2=p^2/3$,
\begin{eqnarray}
 \frac{1}{3}\sum_p \hbar^2 p^2 f_p (1-f_p) &&
= \frac{8 \pi m}{3} \frac{\Omega_0}{(2 \pi)^3} \int E_p \left( -\frac{\partial f_p}{\partial \beta E_p} \right) p^2 dp 
= - \frac{4 \pi m}{3 \beta} \frac{(2 m)^{3/2}}{\hbar^3} \frac{\Omega_0}{(2 \pi)^3} \int \frac{\partial f_p}{\partial  E_p} E_p^{3/2} d E_p 
\nonumber \\ &&  
= \frac{2 \pi m}{ \beta} \frac{(2 m)^{3/2}}{\hbar^3} \frac{\Omega_0}{(2 \pi)^3} \int  f_p E_p^{1/2} d E_p 
= \frac{ m}{ \beta}  \frac{4 \pi \Omega_0}{(2 \pi)^3} \int  f_p p^2 dp = \frac{ m}{ \beta} \sum_p f_p= \frac{ Nm}{ \beta}
\end{eqnarray}
after integration by parts. This is also identical with $\left( \hat P_1, \hat P_1 \right)$ which is the Kubo scalar product (\ref{kubokf}) of the first moment (\ref{Pnu}).

In the collision term, that is the second term on the right hand side of \rf{dynkinequc}, 
we insert the expression \rf{Lei}. The sum over $p'$ is immediately executed and gives $q_E$. 
The first contribution ( from $\delta$ function) as well as to the third contribution (from first principal part)  are considered together and can be transformed by
 $\vec q \to - \vec q$, then $\vec p \to \vec p+\vec q$, 
 so that they coincide with the second and fourth  contributions, respectively.
We find after canceling some common factors 
\begin{eqnarray}
\label{dynkinequc2}
e \hbar^2N \tilde E &&=  - i  \omega \frac{m}{\beta} \hbar^2 N F_1 
-    \sum_{q}  |V_{\rm ei}(q)|^2 q_E^2 
 \sum_p\frac{f_p-f_{p+q}}{\beta (E_{p+q}-E_p)} 
\,\, \frac{i}{\omega +i \epsilon +(E_{p+q}-E_p)/\hbar}F_1\,.
\end{eqnarray}

From \rf{onemoment} we find
\begin{equation}
\nu_{\rm D}(\omega)=-\frac{ \beta}{mN}\sum_{p,q}  q_E^2 |V_{\rm ei}(q)|^2 \frac{f_p-f_{p+q}}{\beta (E_{p+q}-E_p)} 
 \frac{i}{\omega +i \epsilon +(E_{p+q}-E_p)/\hbar}\,.
\end{equation}
We shift $\vec p \to \vec p -\vec q/2$ so that $E_{p+q/2}-E_{p-q/2}= \hbar^2 \vec p \cdot \vec q/m$ and with spin factor 2,
\begin{eqnarray}
\nu_{\rm D}(\omega)=&&\frac{\beta}{mN}\sum_{q} q^2_E |V_{\rm ei}(q)|^2 \frac{m}{\beta \hbar^2 q}\frac{2 \Omega_0}{(2 \pi)^2}
\int_{-\infty}^\infty ds  \frac{1}{s}  \frac{i}{\omega +i \epsilon +\hbar qs/m} \nonumber\\
&&\times \int_0^\infty r dr
\frac{e^{\beta (\hbar^2/2 m) (r^2+s^2+ sq+q^2/4)-\beta \mu}-e^{\beta (\hbar^2/2 m) (r^2+s^2- sq+q^2/4)-\beta \mu}}{(e^{\beta (\hbar^2/2 m) (r^2+s^2+ sq+q^2/4)-\beta \mu}+1)(e^{\beta (\hbar^2/2 m) (r^2+s^2- sq+q^2/4)-\beta \mu}+1) }\,,
\label{D4} 
\end{eqnarray}
where cylindrical coordinates with respect to the $\vec q$ direction have been introduced. $s$ is the component of 
$\vec p$ in $\vec q$ direction, $r$ is the component orthogonal to this axis.
The integral over $r$ can be performed,
\begin{eqnarray}
\label{dynkinequf}
&& \frac{1}{2} \int_0^\infty \, dr^2\,   \frac{1}{e^{\beta (\hbar^2/2 m) (r^2+s^2+ sq+q^2/4)-\beta \mu}+1}  
\frac{1}{e^{-\beta (\hbar^2/2 m) (r^2+s^2- sq+q^2/4)+\beta \mu}+1} 
\nonumber\\ 
=&& \frac{m}{\beta \hbar^2}\frac{1}{e^{\beta (\hbar^2/m)sq}-1} \ln \left[\frac{1+e^{-\beta (\hbar^2/2 m) 
(s-q/2)^2+\beta \mu}}{1+e^{-\beta (\hbar^2/2 m) (s+q/2)^2+\beta \mu}}\right]\,.
\end{eqnarray}
 Furthermore, we neglect the ion correlation so that $S(\vec q) =1$ for the structure factor. Note, that the Born approximation, \rf{D4}, is divergent at zero frequency. As well known, this problem is solved if we go beyond the Born approximation and take into account higher order contributions due to dynamical screening and strong collisions. This was already discussed by Landau and Lifshitz \cite{landau10chap46} and has been shown to be consistent using Green function techniques, see \cite{rerrw00,Reinholz05,Roep98}. In this way, the correct zero frequency limit of the collision frequency is obtained. As shown in Ref.\,\cite{Morales},   alternatively, the Coulomb potential in  \rf{D4} can  be replaced by a statically screened potential, the Debye potential (\ref{Vdeb}), 
 so that  $ |V_{\rm ei}(q)|^2 \approx N V_D^2$.
With $ s=\sqrt{\frac{2m}{\beta \hbar^2}}\frac{x}{y}, q=\sqrt{\frac{8m}{\beta \hbar^2}} y$, expression (\ref{nudegen}) follows.

\section{Renormalization factor and dynamical conductivity}
\label{app4}

We use Rydberg units as introduced at the beginning of Section \ref{subsx1}  and in  \rf{Rydb}.
In LRT, the conductivity \ref{Drude2} within one-moment Born approximation in the non-degenerate limit  \ref{nuclassic} gives ($w=\omega^* \sqrt{\pi n}/T$)
\begin{equation}
\label{sigmaLRT*}
\sigma^*_{\rm LRT}=-\sqrt{\frac{ 2n}{\pi T^3}}\left[ i \omega^*- i \frac{2}{3 \pi} \sqrt{\frac{ n}{ T^3}}
r(w) \int_0^\infty dy \frac{y^4}{(y^2+2 \pi n/T^2)^2} \int_{-\infty}^\infty dx  \frac{1-e^{-4 xy} }{xy (w-xy +i \varepsilon) } e^{-(x-y)^2}  \right]^{-1}\,.
\end{equation}

The renormalization factor $r(w)$ is taken with the first and third moment of the distribution function (i.e. particle current and energy current). According to \rf{2sigma}, generalized force-force correlation functions have to be calculated after decomposition:
$\langle  \dot P_l; \dot P_m \rangle_{\omega+i \epsilon}=\langle  \dot P^{\rm ei}_l; \dot P^{\rm ei}_m \rangle_{\omega+i \epsilon}
+\langle  \dot P^{\rm ee}_l; \dot P^{\rm ee}_m \rangle_{\omega+i \epsilon}$\,.
Considering non-degenerate limit of the  Born approximation again, we have from the electron-ion interaction 
\begin{equation}
\langle  \dot P^{\rm ei}_l; \dot P^{\rm ei}_m \rangle_{\omega+i \epsilon}=i  
\frac{4 }{3\sqrt{\pi}} \frac{N n}{\sqrt{ T}}
 \int_0^\infty dy \frac{y^4}{(y^2+2 \pi n/T^2)^2} \int_{-\infty}^\infty dx  \frac{1-e^{-4 xy} }{xy (w-xy +i \varepsilon) } e^{-(x-y)^2}  \left\{x,y\right\}^{\rm ei}_{lm} 
\end{equation}
where $ \left\{x,y\right\}^{\rm ei}_{11}=1,\,\,\,   \left\{x,y\right\}^{\rm ei}_{31}= 1+3 x^2 +y^2, 
$ and $  \left\{x,y\right\}^{\rm ei}_{33}= 2+10 x^2 +9 x^4+2 y^2+6 x^2 y^2+y^4$.

For the electron-electron interaction we find
\begin{equation}
\langle  \dot P^{\rm ee}_l; \dot P^{\rm ee}_m \rangle_{\omega+i \epsilon}=-i 
\frac{4 }{3\sqrt{2\pi}} \frac{N n}{\sqrt{ T}}
 \int_0^\infty dy \frac{y^4}{(y^2+4 \pi n/T^2)^2} \int_{-\infty}^\infty dx  \frac{1-e^{-4 xy} }{xy (w-xy +i \varepsilon) } e^{-(x-y)^2} 
 \left\{x,y\right\}^{\rm ee}_{lm}
\end{equation}
where due to momentum conservation ($ \dot P^{\rm ee}_1=0$) we have $ \left\{x,y\right\}^{\rm ee}_{11}= \left\{x,y\right\}^{\rm ee}_{31}= 0$ and 
$ \left\{x,y\right\}^{\rm ee}_{33}= 1+(19/4) x^2$.

For the evaluation we use $\frac{1}{xy-w-i \epsilon}= \frac{\cal P}{xy-w} +i \pi \delta (xy - w)$. The $\delta$ function allows to
perform  the integral over $x$ to obtain the real part of the correlation functions $\langle  \dot P_l; \dot P_m \rangle_{\omega+i \epsilon}$. For the imaginary part, we also can perform the $x$ integral after partial fraction decomposition and using
${\cal P} \int_{-\infty}^\infty dx \frac{e^{-x^2}}{x+a}=\pi e^{-a^2} {\rm erfi}(a)$.

In particular, we have for the single moment approximation where $r(w)=1$
\begin{eqnarray} \label{sigmaLRT*1}
\sigma^*_{\rm LRT,1}&&=-\sqrt{\frac{ 2n}{\pi T^3}} \\  \nonumber 
&& \times \left[ i \omega^*- \frac{2}{3w} \sqrt{\frac{ n}{ T^3}}
 \int_0^\infty dy \frac{y^3}{(y^2+\frac{2 \pi n}{T^2})^2}  
 \left\{e^{-(y-\frac{w}{y})^2}-e^{-(y+\frac{w}{y})^2}-2i\left(e^{-(y-\frac{w}{y})^2} {\rm erfi}(y-\frac{w}{y})-e^{-y^2} {\rm erfi}(y)\right) \right\}\right]^{-1}.
\end{eqnarray}

For direct comparison, we give explicitely the dynamical conductivity from  KT (\ref{sigmaKT}) with the energy dependent relaxation time for the Lorentz plasma (\ref{coulomblog})
\begin{equation}
\label{sigmaKT*}
\sigma^*_{\rm KT}=-\frac{8 }{3 \sqrt{\pi}} \sqrt{\frac{ 2n}{\pi T^3}} \frac{1}{T} \int_0^\infty dx \frac{x^4 e^{-x^2/T}}{i \omega^*
- \sqrt{\pi n}\left[ \ln (1+b)-b/(1+b)\right]/x^3}
\end{equation}
with $b=x^2T/(2 \pi n)$.

\begin{acknowledgments}
The authors acknowledge support within the DFG funded Special Research Centre SFB 652. G.R. thanks for the financial support from a 
Research Fellowship of the Johannes Kepler University and the hospitality during his stay at the Johannes Kepler University. We thank J. Adams, M. Winkel, M. Veysman and T. Raitza for fruitful discussions on the presented topic.

\end{acknowledgments}

\end{document}